\documentclass{book}
\usepackage{mypref2}
\usepackage{epsfig}
\newcommand{\be}{\begin{equation}}
\newcommand{\ee}{\end{equation}}
\newcommand{\ba}{\begin{eqnarray}}
\newcommand{\ea}{\end{eqnarray}}

\newcommand{\subs}[1]{{\mbox{\scriptsize #1}}}
\newcommand{\tfrac}[2]{{\textstyle\frac{#1}{#2}}}

\newenvironment{myref}[1]
   {\vspace{0.5cm}\noindent$\bullet$ {\sc #1}:\nopagebreak[3]
    \vspace{0.5cm}
    \begin{list}{}{\setlength{\leftmargin}{0cm}}}
   {\end{list}}

\begin{document}

\title{{\bf THEORY OF FINANCIAL RISK:}\\
{\sc  Basic notions in probability}}
\author{Jean-Philippe Bouchaud\\and\\Marc Potters}
\date{
DRAFT\\
\today\\
\vskip 2cm
{\small
This text corresponds to Chapter 1 of\\
{\em Theory of Financial Risk}\\
Evaluation copy available for free at\\
http://www.science-finance.fr/\\
\vskip 1cm
send comments, typos, etc.\ to\\
book@science-finance.fr\\
}
}
\maketitle
\vfill\eject

\setcounter{page}{5}
\renewcommand{\thepage}{\roman{page}}
\tableofcontents
\setcounter{page}{1}
\renewcommand{\thepage}{\arabic{page}}
\chapter{Probability theory: basic notions}

\begin{quote}
{\sl All epistemologic value of the theory of probability is 
based on this: that large scale random phenomena in their collective
action create strict, non random regularity.}

(Gnedenko et Kolmogorov, Limit Distributions for Sums of Independent 
Random Variables.)
\end{quote}

\section{Introduction}

Randomness stems from our incomplete knowledge of reality, from
the lack of information which forbids a perfect prediction of the
future: randomness arises from complexity, from the fact that causes
are diverse, that tiny perturbations may result in large effects. For
over a century now, Science has abandoned Laplace's deterministic
vision, and has fully accepted the task of deciphering randomness and
inventing adequate tools for its description. The surprise is that,
after all, randomness has many facets and that there are many levels
to uncertainty, but, above all, that a new form of predictability
appears, which is no longer deterministic but {\it statistical}.

Financial markets offer an ideal testing ground for these statistical
ideas: the fact that a large number of participants, with divergent
anticipations and conflicting interests, are simultaneously present in
these markets, leads to an unpredictable behaviour. Moreover, financial
markets are (sometimes strongly) affected by external news -- which
are, both in date and in nature, to a large degree unexpected. The
statistical approach consists in drawing from past observations some
information on the frequency of possible price changes, and in
assuming that these frequencies reflect some intimate mechanism of the
markets themselves, implying that these frequencies will remain stable
in the course of time. For example, the mechanism underlying the
roulette or the game of dice is obviously always the same, and one
expects that the frequency of all possible outcomes will be invariant
in time -- although of course each individual outcome is random.
 
\begin{figure}
\centerline{\hbox{
\epsfig{figure=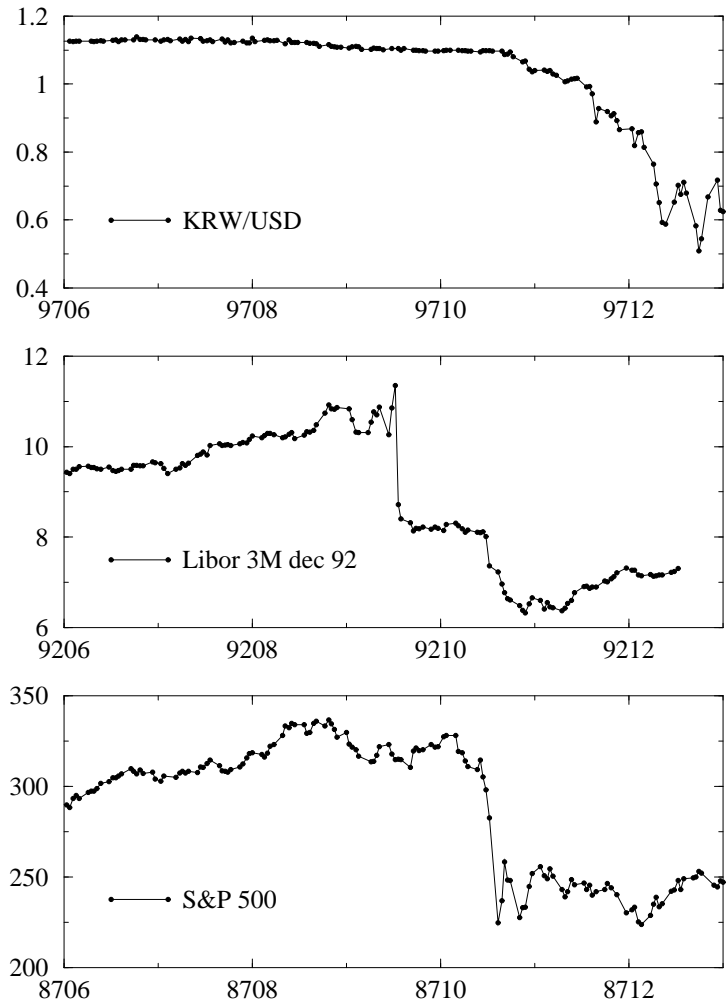,width=9.5cm}
}}
\caption{Three examples of statistically unforseen crashes:
the Korean won against the dollar in 1997 (top), the British 3 month
short term interest rates futures in 1992 (middle), and the S\&P 500
in 1987 (bottom). In the example of the Korean Won, it is particularly
clear that the distribution of price changes before the crisis was
extremely narrow, and could not be extrapolated to anticipate what
happened in the crisis period. }
\label{FigI0}\label{sp1}\label{crash1}
\end{figure}

This `bet' that probabilities are stable (or better, stationary) is
very reasonable in the case of roulette or dice;\footnote{The idea
that Science ultimately amounts to making the best possible guess of
reality is due to R. P. Feynman, (`Seeking New Laws', in {\it The
Character of Physical Laws}, MIT Press, 1967).} it is nevertheless
much less justified in the case of financial markets -- despite the
large number of participants which confer to the system a certain
regularity, at least in the sense of Gnedenko and Kolmogorov.It is
clear, for example, that financial markets do not behave now as they
did thirty years ago: many factors contribute to the evolution of the
way markets behave (development of derivative markets, worldwide and
computer-aided trading, etc.). As will be mentioned in the following,
`young' markets (such as emergent countries markets) and more mature
markets (exchange rate markets, interest rate markets, etc.) behave
quite differently. The statistical approach to financial markets is
based on the idea that whatever evolution takes place, this happens
sufficiently {\it slowly} (on the scale of several years) so that the
observation of the recent past is useful to describe a not too distant
future.  However, even this `weak stability' hypothesis is sometimes
badly in error, in particular in the case of a crisis, which marks a
sudden change of market behaviour. The recent example of some Asian
currencies indexed to the dollar (such as the Korean won or the Thai
baht) is interesting, since the observation of past fluctuations is
clearly of no help to predict the sudden turmoil of 1997 -- see Fig.\
\ref{FigI0}.

Hence, the
statistical description of financial fluctuations is certainly
imperfect. It is nevertheless extremely helpful: in practice, the
`weak stability' hypothesis is in most cases reasonable, at least to
describe {\it risks}.\footnote{The prediction of {\it future returns}
on the basis of past returns is however much less justified.}

In other words, the amplitude of the possible price changes (but not
their sign!) is, to a certain extent, predictable. It is thus rather
important to devise adequate tools, in order to {\it control} (if at
all possible) financial risks. The goal of this first chapter is 
to present a certain number of basic notions in probability theory,
which we shall find useful in the following. Our presentation does not
aim at mathematical rigour, but rather tries to present the key
concepts in an intuitive way, in order to ease their empirical use in
practical applications.

\section{Probabilities}

\subsection{Probability distributions}

Contrarily to the throw of a dice, which can only return an integer
between 1 and 6, the variation of price of a financial \label{asset}
asset\footnote{{\it Asset}\/ is the generic name for a financial
instrument which can be bought or sold, like stocks, currencies, gold,
bonds, etc.} can be arbitrary (we disregard the fact that price
changes cannot actually be smaller than a certain quantity -- a
`tick').\label{tick} In order to describe a
random process $X$ for which the result is a real number, one uses a
probability density $P(x)$, such that the probability that $X$ is
within a small interval of width $dx$ around $X=x$ is equal to $P(x)
dx$. In the following, we shall denote as $P(.)$ the probability
density for the variable appearing as the argument of the function.
This is a potentially ambiguous, but very useful notation.

The probability that $X$ is between $a$ and $b$ is given by the
integral of $P(x)$ between $a$ and $b$, \be {\cal P}
(a<X<b)=\int_a^b P(x)dx.  \ee In the following, the notation ${\cal P}
(.)$ means the probability of a given event, defined by the content of
the parenthesis $(.)$.

The function $P(x)$ is a density; in this sense it depends on the
units used to measure $X$. For example, if $X$ is a length measured in
centimetres, $P(x)$ is a probability density per unit length, i.e.\ per
centimetre. The numerical value of $P(x)$ changes if $X$ is measured
in inches, but the probability that $X$ lies between two specific
values $l_1$ and $l_2$ is of course independent of the chosen
unit. $P(x) dx$ is thus invariant upon a change of unit, i.e.\  under
the change of variable $x \to \gamma x$. More generally, $P(x) dx$ is
invariant upon any (monotonous) change of variable $x \to y(x)$: in
this case, one has $P(x) dx = P(y) dy$.

In order to be a probability density in the usual sense, $P(x)$ must
be non negative ($P(x)\geq 0$ for all $x$) and must be normalised,
that is that the integral of $P(x)$ over the whole range of possible
values for $X$ must be equal to one:
\be\label{E_NORM}
\int_{x_m}^{x_M} P(x) dx =1,
\ee
where $x_m$ (resp.\ $x_M$) is the smallest value \label{loicumulee}
(resp.\ largest) which $X$ can take. In the case where the possible
values of $X$ are not bounded from below, one takes $x_m=-\infty$, and
similarly for $x_M$.  One can actually always assume the bounds to be
$\pm \ \infty$ by setting to zero $P(x)$ in the intervals
$]-\infty,x_m]$ and $[x_M,\infty[$. Later in the text, we shall
often use the symbol $\int$ as a shorthand for $\int_{-\infty}^{+\infty}$.

An equivalent way of describing the distribution of $X$ is to consider
its cumulative distribution ${\cal P}_{<} (x)$, defined as:
\be\label{equiv}
{\cal P}_{<}(x)\equiv {\cal P} (X<x)=\int_{-\infty}^{x} P(x') dx'.
\ee
${\cal P}_{<}(x)$ takes values between zero and one, and is
monotonously increasing with $x$.  Obviously, ${\cal P}_{<}(-\infty)=0$
and ${\cal P}_{<}(+\infty)=1$. Similarly, one defines ${\cal
P}_{>}(x)=1-{\cal P}_{<}(x)$.

\subsection{Typical values and deviations}
\label{S_VALTYP}

\begin{figure}
\centerline{\hbox{
\epsfig{figure=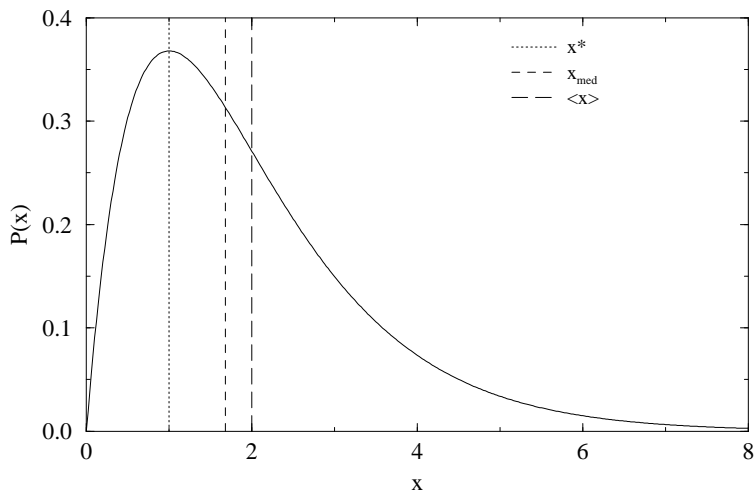,width=10cm}
}}
\caption{The `typical value' of a random variable $X$ drawn according to
a distribution density $P(x)$ can be defined in at least three
different ways: through its mean value $\langle x \rangle$, its most
probable value $x^*$ or its median $x\protect\subs{med}$. In the
general case these three values are distinct. } \label{FigI1}
\end{figure}

It is rather natural to speak about `typical' values of $X$. There are
at least three mathematical definitions of this intuitive notion: the
{\it most probable} value, the {\it median} and the {\it mean}.\label{mean}
\label{median} The
most probable value $x^*$ corresponds to the maximum of the function
$P(x)$; $x^*$ needs not be unique if $P(x)$ has several equivalent
maxima. The median $x_\subs{med}$ is such that the probabilities that
$X$ be greater or less than this particular value are equal. In other
words, ${\cal P}_{<}(x_\subs{med})={\cal
P}_{>}(x_\subs{med})=\frac{1}{2}$.  The mean, or {\it expected
value} of $X$, which we shall note as $m$ or $\langle x \rangle$ in
the following, is the average of all possible values of $X$, weighted
by their corresponding probability:
\be
m\equiv \langle x \rangle=\int x P(x) dx.
\ee
For a unimodal distribution (unique maximum), symmetrical around this
maximum, these three definitions coincide. However, they are in
general different, although often rather close to one another. Figure
\ref{FigI1} shows an example of a non symmetric distribution,
and the relative position of the most probable value, the median and
the mean.

One can then describe the fluctuations of the random variable $X$: if
the random process is repeated several times, one expects the results
to be scattered in a cloud of a certain `width' in the region of
typical values of $X$. This width can be described by the {\it mean
absolute deviation} ({\sc mad}) $E_\subs{abs}$, by the {\it root
mean square} ({\sc rms}\label{rms}) $\sigma$ (or, in financial terms, the {\it
volatility} \label{volat1}), or by the `full width at half maximum' $w_{1/2}$.

The mean absolute deviation from a given reference value is the
average of the distance between the possible values of $X$ and this
reference value,\footnote{One chooses as a reference value the median
for the {\sc mad} and the mean for the {\sc rms}, because for a
fixed distribution $P(x)$, these two quantities minimise,
respectively, the {\sc mad} and the {\sc rms}.}
\be
E_\subs{abs}\equiv\int |x-x_\subs{med}| P(x)dx.
\ee
Similarly, the {\it variance} ($\sigma^2$) is the mean distance squared
to the reference value $m$,
\be\label{E_VAR}
\sigma^2\equiv\langle (x-m)^2 \rangle =\int (x-m)^2 P(x)dx.
\ee
Since the variance has the dimension of $x$ squared, its square
root (the {\sc rms} $\sigma$) gives the order of magnitude of
the fluctuations around $m$.

Finally, the full width at half maximum $w_{1/2}$ is defined (for a
distribution which is symmetrical around its unique maximum $x^*$)
such that $P(x^* \pm \frac{w_{1/2}}{2})=\frac{P(x^*)}{2}$, which
corresponds to the points where the probability density has dropped by
a factor two compared to its maximum value.  One could actually define
this width slightly differently, for example such that the total
probability to find an event outside the interval
$[x^*-\frac{w}{2},x^* +\frac{w}{2}]$ is equal to -- say -- $0.1$.

The pair mean-variance is actually much more popular than the pair
median-{\sc mad}. This comes from the fact that the absolute value is
not an analytic function of its argument, and thus does not possess
the nice properties of the variance, such as additivity under
convolution, which we shall discuss below. However, for the empirical
study of fluctuations, it is sometimes preferable to use the {\sc
mad}; it is more {\it robust} than the variance, that is, less
sensitive to rare extreme events, source of large statistical errors.

\subsection{Moments and characteristic function}
\label{fonctcar}

More generally, one can define higher order {\it moments} of the
distribution $P(x)$ as the average of powers of $X$:
\be\label{E_MOMENT} m_{n}\equiv\langle
x^n\rangle=\int x^n P(x)dx.  
\ee
Accordingly, the mean $m$ is the first moment 
 ($n=1$), while the variance is related to the second moment ($\sigma^2=m_2-m^2$).
The above definition (\ref{E_MOMENT}) is only meaningful if the integral converges, which
requires that $P(x)$ decreases sufficiently rapidly for large $|x|$ (see below).

From a theoretical point of view, the moments are interesting: if they 
exist, their knowledge is often equivalent to the knowledge of the distribution  $P(x)$ 
itself.\footnote{This is not rigourously correct, since one can exhibit examples of different
distribution densities which possess exactly the same moments: see \ref{S_LOGNORMALE} below.}
In practice however, the high order moments are very hard to determine satisfactorily: as $n$ grows, longer and
longer time series are needed to keep a certain level of precision on $m_n$; these high moments are
thus in general not adapted to describe empirical data. 

For many computational purposes, it is convenient to introduce the {\it characteristic
function} of $P(x)$, defined as its Fourier transform:

\be\label{E_FONCAR}
\hat{P}(z) \equiv \int e^{izx} P(x)dx.
\ee
The function $P(x)$ is itself related to its characteristic function through an inverse
Fourier transform:
\be
P(x)=\frac{1}{2\pi}\int e^{-izx} \hat{P}(z)dz.
\ee
Since $P(x)$ is normalised, one always has $\hat{P}(0)=1$. The moments of $P(x)$ 
can be obtained through successive derivatives of the characteristic function at $z=0$, 
\be\label{E_MOMENT2}
m_{n}=(-i)^n\left.\frac{d^n}{dz^n}\hat{P}(z)\right|_{z=0}.
\ee
One finally define the {\it cumulants} $c_n$ of a distribution as the successive derivatives
of the logarithm of its characteristic function:
\be\label{E_CUMULANT}
c_n=(-i)^n\left.\frac{d^n}{dz^n}\log\hat{P}(z)\right|_{z=0}.
\ee
The cumulant \label{cumulant} $c_n$ is a polynomial combination of the moments
$m_p$ with  $p\leq n$. For example $c_2=m_2-m^2=\sigma^2$.
It is often useful to normalise the cumulants by an appropriate power of the 
variance, such that the resulting quantity is dimensionless. One thus define the
{\it normalised cumulants}\/
$\lambda_n$,
\be\label{E_CUMNORM}
\lambda_n\equiv c_n/\sigma^n.
\ee
One often uses the third and fourth normalised cumulants, called the
\label{skewness} {\em skewness} and \label{kurtosis} {\em kurtosis}\/
($\kappa$),\footnote{Note that it is sometimes $\kappa+3$, rather than
$\kappa$ itself, which is called the kurtosis.}
\be\label{E_KURTOSIS}
\lambda_3=\frac{\langle(x-m)^3\rangle}{\sigma^3}
\qquad
\kappa\equiv \lambda_4=\frac{\langle(x-m)^4\rangle}{\sigma^4}-3.
\ee
The above definition of cumulants may look arbitrary, but these
quantities have remarkable properties. For example, as we shall show
in Section \ref{S_ADDITION}, the cumulants simply add when one sums
independent random variables.  Moreover a Gaussian distribution (or
the normal law of Laplace and Gauss) is characterised by the fact that
all cumulants of order larger than two are identically zero. Hence the
cumulants, in particular $\kappa$, can be interpreted as a measure of
the distance between a given distribution $P(x)$ and a
Gaussian.

\subsection{Divergence of moments -- Asymptotic behaviour}

The moments (or cumulants) of a given distribution do not always exist. A necessary condition 
for the $n^{th}$ moment ($m_n$) to exist is that the distribution density $P(x)$ should 
decay faster than  $1/|x|^{n+1}$ for $|x|$ going towards infinity, or else the integral (\ref{E_MOMENT})
would diverge for $|x|$ large. If one restricts to distribution densities behaving asymptotically as
a power law, with an exponent \label{loidepuissance}
$1+\mu$, 
\be\label{E_ASYM}
P(x) \sim 
\frac{\mu A^\mu_{\pm}}{|x|^{1+\mu}}\mbox{ for $x\rightarrow\pm\infty$,}
\ee
then all the moments such that  $n\geq\mu$ are infinite. For example, such a distribution has no
finite variance whenever $\mu\leq 2$.
[Note that, for $P(x)$ to be a normalisable probability distribution, the integral (\ref{E_NORM}) must converge, 
which requires $\mu>0$.]
\begin{technical}
The characteristic function of a distribution having an asymptotic power law behaviour given by
 (\ref{E_ASYM}) is non analytic around $z=0$. The small $z$ expansion contains regular terms of the form
$z^n$ for  $n<\mu$ followed by a non analytic term $|z|^\mu$ (possibly with logarithmic
corrections such as $|z|^\mu\log z$ for integer $\mu$). The derivatives of order larger or equal to
 $\mu$ of the characteristic function thus do not exist at the origin ($z=0$). 
\end{technical}

\section{Some useful distributions}

\subsection{Gaussian distribution}
\label{gaussienne}
The most commonly encountered distributions are the `normal'
laws of Laplace and Gauss, which we shall simply call in the following
Gaussians. Gaussians are ubiquitous: for example, the number of {\it
heads}\/ in a sequence of a thousand coin tosses, the exact number of
oxygen molecules in the room, the height (in inches) of a randomly
selected individual, are all approximately described by a Gaussian
distribution.\footnote{Although, in the above three examples, the
random variable cannot be negative. As we shall discuss below, the
Gaussian description is generally only valid in a certain
neighbourhood of the maximum of the distribution.} The ubiquity of the
Gaussian can be in part traced to the Central Limit Theorem ({\sc
clt}) discussed at length below, which states that a phenomenon
resulting from a large number of small independent causes is
Gaussian. There exists however a large number of cases where the
distribution describing a complex phenomenon is {\it not}\/ Gaussian:
for example, the amplitude of earthquakes, the velocity differences in
a turbulent fluid, the stresses in granular materials, etc., and, as
we shall discuss in next chapter, the price fluctuations of most
financial assets.

A Gaussian of mean  $m$  and root mean square $\sigma$ is defined as:
\be\label{E_NORMALE}
P_G(x)\equiv\frac{1}{\sqrt{2 \pi \sigma^2}}
\exp\left(-\frac{(x-m)^2}{2\sigma^2}\right).
\ee
The median and most probable value are in this case equal to $m$,
while the {\sc mad} (or any other definition of the width) is
proportional to the {\sc rms} (for example,
$E_{\subs{abs}}=\sigma\sqrt{2/\pi}$). For $m=0$, all the odd moments
are zero while the even moments are given by
$m_{2n}=(2n-1)(2n-3)... \sigma^{2n}=(2n-1)!!\ \sigma^{2n}$.

All the cumulants of order greater than two are zero for a Gaussian. 
This can be realised by examining its characteristic function:
\be
\hat{P}_G(z)=\exp\left(-\frac{\sigma^2 z^2}{2}+imz\right).
\ee
Its logarithm is a second order polynomial, for which all derivatives
of order larger than two are zero. In particular, the kurtosis of a
Gaussian variable is zero. As mentioned above, the kurtosis is often
taken as a measure of the distance from a Gaussian distribution. When
$\kappa>0$ ({\em leptokurtic} distributions), the corresponding
distribution density has a marked peak around the mean, and rather
`thick' tails. Conversely, when $\kappa<0$, the distribution density
has a flat top and very thin tails. For example, the uniform
distribution over a certain interval (for which tails are absent) has
a kurtosis $\kappa=-\frac{6}{5}$.

A Gaussian variable is peculiar because `large deviations' are
extremely rare. The quantity $\exp(-x^2/2 \sigma^2)$ decays so fast
for large $x$ that deviations of a few times $\sigma$ are nearly
impossible. For example, a Gaussian variable departs from its most
probable value by more than $2 \sigma$ only 5\% of the times, of more
than $3\sigma$ in 0.2\% of the times, while a fluctuation of
$10\sigma$ has a probability of less than $2 \times 10^{-23}$; in
other words, it never happens.

\subsection{Log-normal distribution}
\label{S_LOGNORMALE}
\label{lognormale}

Another very popular distribution in mathematical finance is the
so-called `log-normal' law.  That $X$ is a log-normal random variable
simply means that $\log X$ is normal, or Gaussian.  Its use in finance
comes from the assumption that the {\it rate of returns}, rather than
the absolute change of prices, are independent random variables. The
increments of the logarithm of the price thus asymptotically sum to a
Gaussian, according to the {\sc clt} detailed below. The log-normal
distribution density is thus defined as:\footnote{A log-normal
distribution has the remarkable property that the knowledge of all its
moments is not sufficient to characterise the corresponding
distribution.  It is indeed easy to show that the following
distribution: ${1 \over \sqrt{2 \pi}} x^{-1} e^{-{1 \over 2} (\log
x)^2} [1 + a \sin(2\pi \log x)]$, for $|a|\leq 1$, has moments which
are independent of the value of $a$, and thus coincide with those of a
log-normal distribution, which corresponds to $a=0$ ([Feller] p.
227).  }
\be\label{E_LOGNORMALE}
P_{LN}(x)\equiv\frac{1}{x\sqrt{2 \pi \sigma^2}}
\exp\left(-\frac{\log^2(x/x_0)}{2\sigma^2}\right),
\ee
the moments of which being: $m_n = x_0^n e^{n^2 \sigma^2/2}$.

In the context of mathematical finance, one often prefers log-normal
to Gaussian distributions for several reasons. As mentioned above, the
existence of a random rate of return, or random interest rate,
naturally leads to log-normal statistics. Furthermore, log-normals
account for the following symmetry in the problem of exchange
rates:\footnote{This symmetry is however not always obvious. The
dollar, for example, plays a special role. This symmetry can only be
expected between currencies of similar strength.} if $x$ is the rate
of currency A in terms of currency B, then obviously, $1/x$ is the
rate of currency $B$ in terms of A. Under this transformation, $\log
x$ becomes $-\log x$ and the description in terms of a log-normal
distribution (or in terms of any other even function of $\log x$) is
independent of the reference currency. One often hears the following
argument in favour of log-normals: since the price of an asset
cannot be negative, its statistics cannot be Gaussian since the latter
admits in principle negative values, while a log-normal excludes
them by construction.  This is however a red-herring argument, since
the description of the fluctuations of the price of a financial asset
in terms of Gaussian or log-normal statistics is in any case an {\it
approximation} which is only be valid in a certain range. As we shall
discuss at length below, these approximations are totally unadapted to
describe extreme risks. Furthermore, even if a price drop of more than
$100\%$ is in principle possible for a Gaussian process,\footnote{In
the rather extreme case of a $20\%$ annual volatility and a zero
annual return, the probability for the price to become negative after
a year in a Gaussian description is less than one out of three
million} the error caused by neglecting such an event is much smaller
than that induced by the use of either of these two distributions
(Gaussian or log-normal). In order to illustrate this point more
clearly, consider the probability of observing $n$ times `heads' in a
series of $N$ coin tosses, which is exactly equal to $2^{-N}
C_N^n$. It is also well known that in the neighbourhood of $N/2$,
$2^{-N} C_N^n $ is very accurately approximated by a Gaussian of
variance $N/4$; this is however not contradictory with the fact that
$n \geq 0$ by construction!

\begin{figure}
\centerline{\hbox{
\epsfig{figure=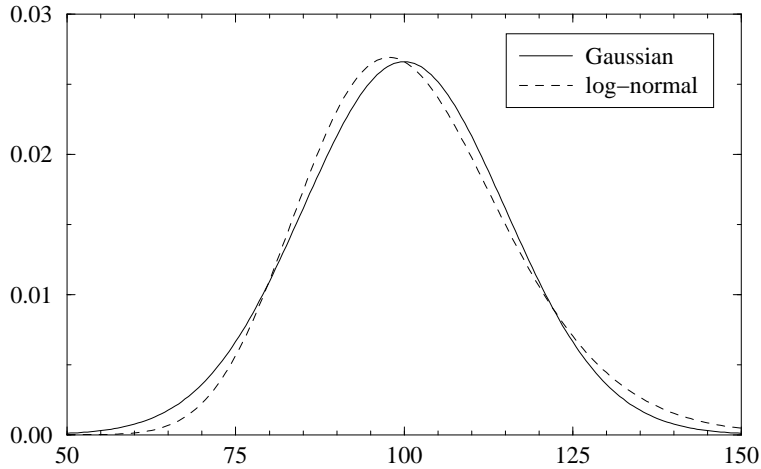,width=10cm}
}}
\caption{Comparison between a Gaussian (thick line) and a log-normal (dashed line), with
$m=x_0=100$ and $\sigma$ equal to $15$ and $15\%$ respectively. The difference between the two curves
shows up in the tails.}\label{FigI2}
\end{figure}

Finally, let us note that for moderate volatilities (up to say
$20\%$), the two distributions (Gaussian and log-normal) look rather
alike, specially in the `body' of the distribution (Fig.\
\ref{FigI2}). As for the tails, we shall see below that Gaussians
substantially underestimate their weight, while the log-normal
predicts that large positive jumps are more frequent than large
negative jumps. This is at variance with empirical observation: the
distributions of absolute stock price changes are rather symmetrical;
if anything, large negative draw-downs are more frequent than large
positive draw-ups.

\subsection{L\'evy distributions and Paretian tails}
\label{loidelevy}
\label{queues}

L\'evy distributions (noted $L_\mu(x)$ below) appear naturally in the
context of the {\sc clt} (see below), because of their stability
property under addition (a property shared by Gaussians).  The tails
of L\'evy distributions are however much `fatter' than those of
Gaussians, and are thus useful to describe multiscale phenomena (i.e.\
when both very large and very small values of a quantity can commonly
be observed -- such as personal income, size of pension funds,
amplitude of earthquakes or other natural catastrophes, etc.). These
distributions were introduced in the fifties and sixties by Mandelbrot
(following Pareto) to describe personal income and the price changes
of some financial assets, in particular the price of cotton
[Mandelbrot]. An important constitutive property of these L\'evy
distributions is their power-law behaviour for large arguments, often
called `Pareto tails':
\be \label{E_ASYM2}
L_\mu(x) \sim 
\frac{\mu A_{\pm}^\mu}{|x|^{1+\mu}} \mbox{ for $x\rightarrow\pm\infty$},
\ee 
where $0 <\mu < 2$ is a certain exponent (often called $\alpha$), and $A_{\pm}^\mu$
two constants which we call  {\it tail amplitudes}, or {\it scale parameters}: $A_\pm$ indeed
gives the order of magnitude of the large \label{amplitude}
(positive or negative)  fluctuations of $x$. For instance, the probability to draw a number larger than $x$
decreases as ${\cal P}_>(x)=(A_+/x)^\mu$ for large positive $x$.

One can of course in principle observe Pareto tails with $\mu \geq 2$,
however, those tails do not correspond to the asymptotic behaviour of
a L\'evy distribution.

In full generality, L\'evy distributions are characterised by an {\it
asymmetry parameter} defined as
$\beta\equiv(A^\mu_{+}-A^\mu_{-})/(A^\mu_{+}+A^\mu_{-})$, which
measures the relative weight of the positive and negative tails. We
shall mostly focus in the following on the symmetric case
$\beta=0$. The fully asymmetric case ($\beta=1$) is also useful to
describe strictly positive random variables, such as, for example, the
time during which the price of an asset remains below a certain value,
etc.

An important consequence of (\ref{E_ASYM}) with $\mu \leq 2$ is that
the variance of a L\'evy distribution is formally infinite: the
probability density does not decay fast enough for the integral
(\ref{E_VAR}) to converge.  In the case $\mu\leq 1$, the distribution
density decays so slowly that even the mean, or the {\sc mad}, fail to
exist.\footnote{The median and the most probable value however still
exist. For a symmetric L\'evy distribution, the most probable value
defines the so-called `localisation' parameter $m$.}  The scale of the
fluctuations, defined by the width of the distribution, is always set
by $A=A_+=A_-$.

\begin{figure}
\centerline{\hbox{
\epsfig{figure=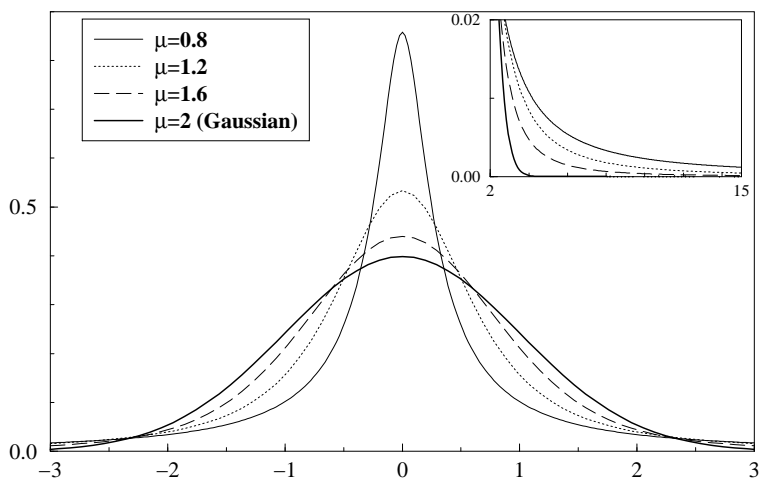,width=10cm}
}}
\caption{Shape of the symmetric L\'evy distributions with $\mu=0.8,\ 1.2,\ 1.6$ and $2$ 
(this last value actually corresponds to a Gaussian). The smaller $\mu$, the sharper the
`body' of the distribution, and the fatter the tails, as illustrated in the inset.}
\label{FigI3} \end{figure}

There is unfortunately no simple analytical expression for symmetric L\'evy distributions
$L_{\mu}(x)$,
except for  $\mu=1$, which corresponds to a Cauchy distribution (or `Lorentzian'):  
\be
L_{1}(x)=\frac{A}{x^{2}+\pi^{2}A^2}.
\ee
However, the characteristic function of a symmetric L\'evy
distribution is rather simple, and reads:
\be
\hat{L}_{\mu}(z)=\exp\left(-a_{\mu}|z|^{\mu}\right),\label{E_FLEVY}
\ee
where $a_{\mu}$ is a certain constant, proportional to the tail parameter $A^\mu$.\footnote
{For example, when $1 < \mu < 2$, $A^\mu = \mu 
\Gamma(\mu-1) \sin(\pi \mu/2) a_\mu/\pi$.} It is thus clear that in the limit $\mu=2$, 
one recovers the definition of a Gaussian. When $\mu$ decreases from $2$, the distribution becomes
more and more sharply peaked around the origin and fatter in its tails, while `intermediate' events
loose weight (Fig.\ \ref{FigI3}). These distributions thus describe  
`intermittent' phenomena, very often small, sometimes gigantic. 

Note finally that Eq.\ (\ref{E_FLEVY}) does not define a probability distribution when $\mu > 2$,
 because its
inverse Fourier transform is not everywhere positive.

\begin{technical}

In the case $\beta\neq 0$, one would have:
\be
\hat{L}_{\mu}^{\beta}(z)=\exp\left[-a_{\mu}|z|^{\mu}\left(1+i\beta\tan(\mu\pi/2)
\frac{z}{|z|}\right)\right] \quad (\mu \neq 1).
\ee
\end{technical}

It is important to notice that while the leading asymptotic term for large $x$ is given
 by Eq.\ (\ref{E_ASYM2}), there are subleading terms which can be important for finite $x$. 
The full asymptotic series actually reads:
\be
L_\mu(x) = \sum_{n=1}^\infty \frac{(-)^{n+1}}{\pi n!} \frac{a_\mu^n}{x^{1+n\mu}} \Gamma(1+n\mu)
\sin(\pi \mu n/2)
\ee
The presence of the subleading terms may lead to a bad empirical estimate of the exponent $\mu$ based
on a fit of the tail of the distribution. In particular, the `apparent' exponent which describes
the function  $L_\mu$ for finite $x$ is larger than $\mu$, and decreases towards $\mu$ for $x \to
\infty$, but more and more slowly as $\mu$ gets nearer to the Gaussian value $\mu=2$, for which the power-law
 tails no longer exist. Note however that one also often observes empirically the opposite behaviour, i.e.\ 
an apparent Pareto exponent which {\it grows} with $x$. This arises
when the Pareto 
distribution (\ref{E_ASYM2}) is only valid in an intermediate regime $x \ll 1/\alpha$,
beyond which the distribution decays exponentially, say as $\exp (- \alpha x)$. The Pareto tail
is then `truncated' for large values of $x$, and this leads to an effective $\mu$ which grows with $x$.

An interesting generalisation of the L\'evy distributions which
accounts for this exponential cut-off is given by the `truncated
L\'evy distributions' ({\sc tld}), which will be of much use in the
following. A simple way to alter the characteristic function
(\ref{E_FLEVY}) to account for an exponential cut-off for large
arguments is to set:\footnote{See I.  Koponen, ``Analytic approach to
the problem of convergence to truncated L\'evy flights towards the
Gaussian stochastic process,'' {\it Physical Review E}, {\bf 52},
1197, (1995).}\label{loidelevytronquee}
\be
\hat{L}_{\mu}^{(t)}(z)=\exp\left[-a_{\mu}
\frac{(\alpha^2 + z^2)^{\frac{\mu}{2}}
\cos\left(\mu \mbox{arctan}(|z|/\alpha)\right)
-\alpha^\mu}{\cos (\pi \mu/2)}\right],
\label{E_FLEVYT} 
\ee
for $1 \leq \mu \leq 2$. The above form reduces to  (\ref{E_FLEVY}) for  $\alpha=0$. Note that the argument in the exponential can also be written
as:
\be\label{loidelevytronqueebis}
\frac{a_\mu}{2 \cos (\pi \mu/2)} \left[(\alpha+iz)^\mu + (\alpha-iz)^\mu -2\alpha^\mu\right].
\ee
\subsubsection{Exponential tail: a limiting case}

\begin{technical}
Very often in the following, we shall notice that in the formal limit $\mu \to \infty$,
the power-law tail becomes an exponential tail, if the tail parameter is simultaneously
scaled as $A^\mu =
(\mu/\alpha)^\mu$.  Qualitatively, this can be understood as follows: consider a 
probability distribution restricted to positive $x$, which decays as a power-law for large $x$, defined as:
\be 
{\cal P}_>(x)= \frac{A^\mu}{(A + x)^\mu}.
\ee
This shape is obviously compatible with  (\ref{E_ASYM2}), and is such that 
${\cal P}_>(x=0)=1$. If $A = (\mu/\alpha)$, one then finds:
\be 
{\cal P}_>(x)= \frac{1}{\left(1 + \frac{\alpha x}{\mu}\right)^\mu} 
\mathop{\longrightarrow}_{\mu \to \infty} \exp(-\alpha x).
\ee

\end{technical}

\subsection{Other distributions (*)}
\begin{figure}
\centerline{\hbox{
\epsfig{figure=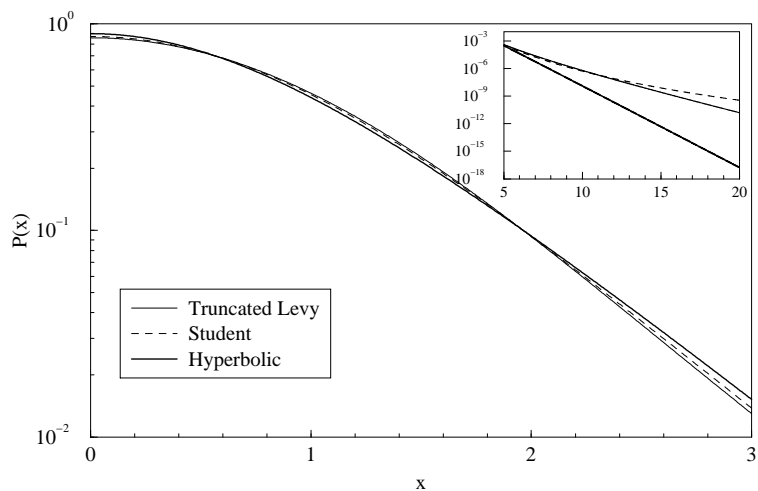,width=10cm}
}}
\caption{Probability density for the truncated L\'evy ($\mu=3/2$), Student and hyperbolic distributions.
All three have two free parameters which were fixed to have unit
variance and kurtosis. The inset shows a blow-up of the tails where
one can see that the Student distribution has tails similar (but slightly
thicker) to that of the truncated L\'evy.} \label{FigI3bis}
\end{figure}
There are obviously a very large number of other statistical distributions useful to
describe random phenomena. Let us cite a few, which often appear in a financial context:

\begin{itemize}
\item The discrete Poisson distribution: consider a set of points randomly
scattered on the real axis, with a certain density  $\omega$ (e.g.\ the 
times when the price of an asset changes). The number of points $n$
in an arbitrary interval of length $\ell$ is distributed according to the Poisson distribution:\label{posson}
\be
P(n) \equiv \frac{(\omega \ell)^n}{n!} \exp(-\omega \ell).\label{poisson}
\ee

\item The hyperbolic distribution, which interpolates between a Gaussian
`body' and exponential tails:\label{hyperbolic}
\be\label{E_HYPERBOLIQUE}
P_H(x) \equiv \frac{1}{2x_0 K_1(\alpha x_0)} \exp -[\alpha \sqrt{x_0^2 + x^2}],
\ee
where the normalisation $K_1(\alpha x_0)$ is a modified Bessel function of the second kind.
For $x$ small compared to $x_0$,  $P_H(x)$
behaves as a Gaussian while its asymptotic behaviour for $x \gg x_0$ is fatter and reads $\exp -\alpha |x|$.

From the characteristic function
\be
{\hat P}_H(z)=\frac{\alpha x_0 K_1(x_0 \sqrt{1+\alpha z})}{K_1(\alpha x_0)\sqrt{1+\alpha z}},
\ee
we can compute the variance 
\be
\sigma^2=\frac{x_0 K_2(\alpha x_0)}{\alpha K_1(\alpha x_0)},
\ee
and kurtosis
\be
\kappa=3\left(\frac{K_2(\alpha x_0)}{K_1(\alpha x_0)}\right)^2+
\frac{12}{\alpha x_0}\frac{K_2(\alpha x_0)}{K_1(\alpha x_0)}-3.
\ee
Note that the kurtosis of the hyperbolic distribution is always between 
zero and three.
In the case  $x_0=0$, one finds the symmetric exponential distribution:\label{exponentialdistribution}
\be
P_E(x) = \frac{\alpha}{2} \exp - \alpha|x|,
\ee
with even moments $m_{2n} = 2n!\,\alpha^{-2n}$, which gives $\sigma^2=2\alpha^{-2}$
and $\kappa=3$. Its characteristic function
reads: ${\hat P}_E(z)={\alpha^2}/(\alpha^2+z^2)$.

\item The Student distribution, which also has power-law tails:\label{student}
\be\label{E_STUDENT}
P_S(x) \equiv \frac{1}{\sqrt{\pi}}\frac{\Gamma((1+\mu)/2)}{\Gamma(\mu/2)}\frac{a^\mu}{(a^2+x^2)^{(1+\mu)/2}},
\ee
which coincides with the Cauchy distribution for $\mu=1$, and tends towards a Gaussian in the limit
$\mu \to \infty$, provided that $a^2$ is scaled as $\mu$. The even moments of the Student distribution
read:  $m_{2n} =(2n-1)!!\Gamma(\mu/2-n)/\Gamma(\mu/2)\,(a^2/2)^{n}$, provided $2n < \mu$; and are infinite
otherwise. One can check that in the limit $\mu \to \infty$, the above expression gives back the
moments of a Gaussian: $m_{2n} =(2n-1)!!\,\sigma^{2n}$. Figure \ref{FigI3bis} shows a plot of the Student
distribution with $\kappa=1$, corresponding to $\mu=10$.
\end{itemize}

\section{Maximum of random variables -- Statistics of extremes}
\label{statistiquedextremes}

If one observes a series of $N$ independent realizations of the same
random phenomenon, a question which naturally arises, in particular
when one is concerned about risk control, is to determine the order of
magnitude of the {\it maximum} observed value of the random variable
(which can be the price drop of a financial asset, or the water level
of a flooding river, etc.).  For example, in Chapter 3, the so-called
`Value-at-Risk' (VaR) on a typical time horizon will be defined as the
possible maximum loss over that period (within a certain confidence
level).

The law of large numbers tells us that an event which has a
probability $p$ of occurrence appears on average $Np$ times on a
series of $N$ observations. One thus expects to observe events which
have a probability of at least $1/N$. It would be surprising to
encounter an event which has a probability much smaller than
$1/N$. The order of magnitude of the largest event observed in a
series of $N$ independent identically distributed ({\sc iid}) random
variables is thus given by:\label{iid1}

\be\label{E_MAXMAX}
{\cal P}_{>}(\Lambda_{\max})=1/N. \ee 

More precisely, the full probability distribution of the maximum value
$x_{\max}=\max_{i=1,N}\{x_i\}$, is relatively easy to characterise;
this will justify the above simple criterion (\ref{E_MAXMAX}).  The
cumulative distribution ${\cal P}(x_{\max}<\Lambda)$ is obtained by
noticing that if the maximum of all $x_{i}$'s is smaller than
$\Lambda$, all of the $x_{i}$'s must be smaller than $\Lambda$. If the
random variables are {\sc iid}, one finds:
\be\label{E_PMAX}
{\cal P}(x_{\max}<\Lambda)=\left[{\cal P}_{<}(\Lambda)\right]^{N}.
\ee
Note that this result is general, and does not rely on a specific choice for  $P(x)$. When
 $\Lambda$ is large, it is useful to use the following approximation:

\be\label{E_PMAX2}
{\cal P}(x_{\max}<\Lambda)=\left[1-{\cal P}_{>}(\Lambda)\right]^{N}
\simeq e^{-N{\cal P}_{>}(\Lambda)}.
\ee
Since we now have a simple formula for the distribution of $x_{\max}$,
one can invert it in order to obtain, for example, the median value of
the maximum, noted $\Lambda_{\subs{med}}$, such that ${\cal
P}(x_{\max}<\Lambda_\subs{med})=1/2$:
\be\label{MAX_MEDIANE}
{\cal P}_{>}(\Lambda_\subs{med})=1-\left(\tfrac{1}{2}\right)^{1/N}
\simeq \frac{\log 2}{N}.
\ee
More generally, the value $\Lambda_{p}$ which is greater than  $x_{\max}$ with probability $p$ is given
by
\be\label{MAX_NIVEAU_P}
{\cal P}_{>}(\Lambda_{p})\simeq -\frac{\log p}{N}. 
\ee
The quantity $\Lambda_{\max}$ defined by Eq.\ (\ref{E_MAXMAX}) above is
thus such that $p=1/e \simeq 0.37$.  The probability that $x_{\max}$
is even {\it larger} than $\Lambda_{\max}$ is thus 63\%.  As we shall
now show, $\Lambda_{\max}$ also corresponds, in many cases, to the
{\it most probable value} of $x_{\max}$.

Equation (\ref{MAX_NIVEAU_P}) will be very useful in Chapter 3 to
estimate a maximal potential loss within a certain confidence
level. For example, the largest daily loss $\Lambda$ expected next
year, with 95\% confidence, is defined such that ${\cal
P}_{<}(-\Lambda)=-\log(0.95)/250$, where ${\cal P}_{<}$ is the
cumulative distribution of daily price changes, and $250$ is the
number of market days per year.

Interestingly, the distribution of $x_{\max}$ only depends, when $N$
is large, on the asymptotic behaviour of the distribution of $x$,
$P(x)$, when $x \to \infty$. For example, if $P(x)$ behaves as an
exponential when $x \to \infty$, or more precisely if ${\cal
P}_{>}(x)\sim \exp(-\alpha x)$, one finds:
\be
\Lambda_{\max}= \frac{\log N}{\alpha},
\ee
which grows very slowly with $N$.\footnote{For example, for a
symmetric exponential distribution $P(x)=\exp(-|x|)/2$, the median
value of the maximum of $N=10\, 000$ variables is only
$6.3$.}\label{exponential1} Setting $x_{\max} =\Lambda_{\max} +
\frac{u}{\alpha}$, one finds that the deviation $u$ around
$\Lambda_{\max}$ is distributed according to the Gumbel distribution:
\be P(u) = e^{-e^{-u}} e^{-u}. \label{E_PU}
\ee
The most probable value of this distribution is $u=0$. This shows that
$\Lambda_{\max}$ is the most probable value of $x_{\max}$. The result
(\ref{E_PU}) is actually much more general, and is valid as soon as
$P(x)$ decreases more rapidly than any power-law for $x \to \infty$:
the deviation between $\Lambda_{\max}$ (defined as (\ref{E_MAXMAX}))
and $x_{\max}$ is always distributed according to
the Gumbel law (\ref{E_PU}), up to a scaling factor in the definition of $u$.

\begin{figure}
\centerline{\hbox{
\epsfig{figure=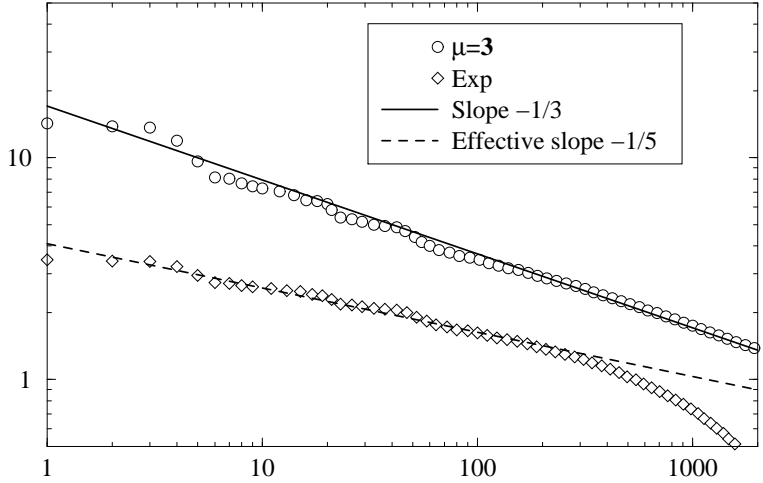,width=10cm}
}}
\caption{Amplitude versus rank plots. One plots the value of the
$n^{th}$ variable $\Lambda[n]$ as a function of its rank $n$.  If
$P(x)$ behaves asymptotically as a power-law, one obtains a straight
line in log-log coordinates, with a slope equal to $-1/\mu$. For an
exponential distribution, one observes an effective slope which is
smaller and smaller as $N/n$ tends to infinity. The points correspond
to synthetic time series of length $5\,000$, drawn according to a
power law with $\mu=3$, or according to an exponential. Note that if
the axis $x$ and $y$ are interchanged, then according to Eq.\
(\protect\ref{rank}), one obtains an estimate of the cumulative
distribution, ${\cal P}_>$. }
\label{FigI4}
\end{figure}
The situation is radically different if  $P(x)$ decreases as a power law,  cf.\ (\ref{E_ASYM}).
In this case, 
\be
{\cal P}_{>}(x)\simeq \frac{A^\mu_{+}}{x^{\mu}}, 
\ee
and the typical value of the maximum is given by:
\be\label{E_MAXPOWER}
\Lambda_{\max} = A_+ N^{\frac{1}{\mu}}. 
\ee
Numerically, for a distribution with  $\mu=3/2$ and a scale factor $A_+ = 1$, the 
largest of  $N=10\, 000$ variables is on the order of 450, while for $\mu=1/2$ it is 
one hundred million! The complete distribution of the maximum, called the Fr\'echet distribution, is given by:
\be
P(u) = \frac{\mu}{u^{1+\mu}} e^{-1/u^\mu}\qquad
u=\frac{x_{\max}}{A_+ N^{\frac{1}{\mu}}}.
\ee
Its asymptotic behaviour for $u \to \infty$ is still a power law of
exponent $1+\mu$.  Said differently, both power-law tails and
exponential tails {\it are stable with respect to the `max'
operation}.\footnote{A third class of laws stable under `max' concerns
random variables which are bounded from above -- i.e.\ such that
$P(x)=0$ for $x>x_M$, with $x_M$ finite. This leads to the Weibull
distributions, which we will not consider further in this book.} The
most probable value $x_{\max}$ is now equal to $(\mu/1+\mu)^{1/\mu}
\Lambda_{\max}$. As mentioned above, the limit 
$\mu \to \infty$ formally  corresponds to an exponential distribution. In this limit, one
indeed recovers $\Lambda_{\max}$ as the most probable value.

\begin{technical}
Equation (\ref{E_MAXPOWER}) allows to discuss intuitively the
divergence of the mean value for $\mu\leq 1$ and of the variance for
$\mu\leq 2$. If the mean value exists, the sum of $N$ random variables
is typically equal to $Nm$, where $m$ is the mean (see also below).
But when $\mu<1$, the largest encountered value of $X$ is on the order
of $N^{1/\mu}\gg N$, and would thus be larger than the entire
sum. Similarly, as discussed below, when the variance exists, the {\sc
rms} of the sum is equal to $\sigma\sqrt{N}$. But for $\mu <2$,
$x_{\max}$ grows faster than $\sqrt{N}$.
\end{technical}

More generally, one can rank the random variables $x_i$ in decreasing
order, and ask for an estimate of the $n^{th}$ encountered value,
noted $\Lambda[n]$ below. (In particular, $\Lambda[1]=x_{\max}$). The
distribution $P_n$ of $\Lambda[n]$ can be obtained in full generality
as:
\be
P_n(\Lambda[n])=C_N^n \ P(x=\Lambda[n]) \ ({\cal P}(x>\Lambda[n])^{n-1}
({\cal P}(x<\Lambda[n])^{N-n}.
\ee
The previous expression means that one has to choose $n$ variables among $N$
as the $n$ largest ones, and then assign the corresponding probabilities 
to the configuration where $n-1$ of them are larger than $\Lambda[n]$ and 
$N-n$ are smaller than $\Lambda[n]$. One can study the position $\Lambda^*[n]$
of the
maximum of $P_n$, and also the width of $P_n$, defined from the second
derivative of $\log P_n$ calculated at $\Lambda^*[n]$. The calculation simplifies
in the limit where $N \to \infty$, $n \to \infty$, with the ratio $n/N$ fixed.
In this limit, one finds a relation which generalises (\ref{E_MAXMAX}):
\be\label{rank}
{\cal P}_{>}(\Lambda^*[n])=n/N. 
\ee 
The width $w_n$ of the distribution is found to be given by:
\be
w_n = \frac{1}{\sqrt{N}} \frac{\sqrt{1-(n/N)^2}}{P(x=\Lambda^*[n])},
\ee
which shows that in the limit $N \to \infty$, the value of the $n^{th}$
variable is more and more sharply peaked around its most probable value
$\Lambda^*[n]$, given by (\ref{rank}). 

In the case of an exponential tail, one finds that $\Lambda[{n}]^* \simeq
{\log(\frac{N}{n})}/{\alpha}$; while in the case of power-law tails, one rather obtains:
\be\label{rank2}
\Lambda^*[n] \simeq
A_+\left(\frac{N}{n}\right)^{\frac{1}{\mu}}.
\ee
This last equation shows that, for power-law variables, the encountered values are 
hierarchically organised: for example, the ratio of the largest value  $x_{\max} \equiv \Lambda[1]$ to the second largest $\Lambda[2]$ is on the order of 
$2^{1/\mu}$, which becomes larger and larger as  $\mu$ decreases, and conversely tends
to one when $\mu 
\to \infty$. \label{historang} 

The property (\ref{rank2}) is very useful to identify empirically the nature of the tails 
of a probability distribution. One sorts in decreasing order the set of observed values
 $\{x_1,x_2,..,x_N\}$ and one simply draws $\Lambda[n]$ as a function of $n$. If the 
variables are power-law distributed, this graph should be a straight line in log-log plot, 
with a slope $-1/\mu$, as given by (\ref{rank2}) (Fig.\ \ref{FigI4}). On the same
figure, we have shown the result obtained for exponentially distributed variables. On this
diagram, one observes an approximately straight line, but with an effective slope 
which varies with the total number of points $N$: the slope is less and less as  $N/n$ 
grows larger. In this sense, the formal remark made above, that an exponential distribution
could be seen as a power law with $\mu \to \infty$, becomes somewhat more concrete.  Note that if the axis $x$ and $y$ of Fig.\ \ref{FigI4} are interchanged, then according 
to Eq.\ (\ref{rank}), one obtains 
an estimate of the cumulative distribution, ${\cal P}_>$.

\begin{technical}

Let us finally note another property of power laws, potentially interesting for their 
empirical determination. If one computes the average value of $x$ conditioned to
a certain minimum value $\Lambda$:
\be
\langle x \rangle_\Lambda = \frac{\int_\Lambda^\infty dx \ x \ P(x)}
{\int_\Lambda^\infty dx \ P(x)},
\ee
then, if $P(x)$ decreases as in (\ref{E_ASYM}), one finds, for $\Lambda \to \infty$,
\be
\langle x \rangle_\Lambda = \frac{\mu}{\mu-1} \Lambda,
\ee
independently of the tail amplitude $A_+^\mu$.\footnote{This means
that $\mu$ can be determined by a one parameter fit only.}  The
average $\langle x \rangle_\Lambda$ is thus always of order of
$\Lambda$ itself, with a proportionality factor which diverges as $\mu
\to 1$.

\end{technical}

\section{Sums of random variables}
\label{S_ADDITION}
In order to describe the statistics of future prices of a financial asset, one {\it a priori}
needs a distribution density for all possible time intervals, corresponding to different trading
time horizons. For example, the distribution of five minutes price fluctuations is different 
from the one describing daily fluctuations, itself different for the weekly, monthly, etc. 
variations. But in the case where the fluctuations are independent and identically distributed
({\sc iid}) -- an assumption which is however not always justified, see  \ref{S_NONSTAT} and  2.4,
it is possible to reconstruct the distributions corresponding to different time scales from
the knowledge of that describing short time scales only. In this context, Gaussians and 
L\'evy distributions play a special role, because they are stable: if the short time scale 
distribution is a stable law, then the fluctuations on all time scales are described by the 
same stable law -- only the parameters of the stable law must be changed (in particular its 
width). More generally, if one sums {\sc iid} variables, then, independently of the short time
 distribution, the law describing 
long times converges towards one of the stable laws: this is the content of the `central limit
theorem' ({\sc clt}). In practice, however, this convergence can be very slow and thus of limited
interest, in particular if one is concerned about short time scales.

\subsection{Convolutions}
\label{convolution}
What is the distribution of the sum of two independent random variable? This sum
can for example represent the variation of price of an asset between today and the day after
tomorrow ($X$), which is the sum of the increment between today and tomorrow ($X_1$) and 
between tomorrow and the day after tomorrow ($X_{2}$),  both assumed to be random and independent. 

Let us thus consider $X=X_1+X_2$ where  $X_1$ and $X_2$ are two random variables, independent, 
and distributed according to $P_1(x_1)$ and  $P_2(x_2)$, respectively. The probability 
that $X$ is equal to $x$ (within $dx$) is given by the sum over all possibilities of obtaining 
$X=x$ (that is all combinations of $X_{1}=x_{1}$ and  $X_{2}=x_{2}$ such that 
$x_{1}+x_{2}=x$), weighted by their respective probabilities. The variables $X_{1}$ and $X_{2}$ 
being independent, the joint probability that $X_{1}=x_{1}$ and 
$X_{2}=x-x_{1}$ is equal to $P_1(x_1) P_2(x-x_1)$, from which one obtains:
\be\label{E_CONV}
P(x,N=2)=\int dx' P_1(x')P_2(x-x') .
\ee
This equation defines the convolution between  $P_1(x)$ and  
$P_2(x)$, which we shall write $P=P_1\star P_2$.  The generalisation to the
sum of $N$ independent random variables is immediate. If  $X=X_1+X_2+...+X_N$ with $X_i$
distributed according to $P_i(x_i)$, the distribution of $X$ is obtained as:
{\small
\be\label{E_NCONV}
P(x,N)=\int \prod_{i=1}^{N-1}dx_i'
     P_1(x_1')... P_{N-1}(x_{N-1}')P_N(x-x_1'-...-x_{N-1}').
\ee
} One thus understands how powerful is the hypothesis that the
increments are {\sc iid}, i.e., that $P_1=P_2=..=P_N$. Indeed,
according to this hypothesis, one only needs to know the distribution
of increments over a unit time interval to reconstruct that of
increments over an interval of length $N$: it is simply obtained by
convoluting the elementary distribution $N$ times with itself.

\begin{technical}
The analytical or numerical manipulations of Eqs.\ (\ref{E_CONV}) and
(\ref{E_NCONV}) are much eased by the use of Fourier transforms, for which convolutions 
become simple products. The equation
$P(x,N=2)=[P_1\star P_2](x)$, reads in Fourier space: 
\be
\hat{P}(z,N=2)=\int dx e^{iz(x-x'+x')} \int dx' P_1(x')P_2(x-x') \equiv \hat{P}_1(z)\hat{P}_2(z).
\ee
In order to obtain the $N^{th}$ convolution of a function with itself, one should raise
its characteristic function to the power $N$, and then take its inverse Fourier transform.
\end{technical}

\subsection{Additivity of cumulants and of tail amplitudes}
\label{S_ADDCUM}
It is clear that the mean of the sum of two random variables (independent or not) is
equal to the sum of the individual means. The mean is thus additive under convolution.
Similarly, if the random variables are independent, one can show that their variances
(if they are well defined) also add simply. More generally, all the cumulants ($c_n$)
of two independent distributions simply add. This follows from the fact that since the 
characteristic functions multiply, their logarithm add. The additivity of cumulants 
is then a simple consequence of the linearity of derivation. 

The cumulants of a given law convoluted $N$ times with itself thus follow the simple
rule  $c_{n,N} = N c_{n,1}$ where the  $\{c_{n,1}\}$ are the cumulants of the elementary
distribution $P_1$. Since the cumulant $c_n$ has the dimension of  $X$ to the power 
$n$, its relative importance is best measured in terms of the normalised cumulants:
\be
\lambda^{N}_n
\equiv \frac{c_{n,N}}{(c_{2,N})^{\frac{n}{2}}}=
\frac{c_{n,1}}{ c_{2,1}}N^{1-n/2}.
\ee
The normalised cumulants thus decay with $N$ for $n>2$; the higher the cumulant, the
faster the decay: $\lambda_n^{N}\propto N^{1-n/2}$. The kurtosis $\kappa$, defined above as the fourth
normalised cumulant, thus decreases as $1/N$. This is basically the content of the {\sc clt}:
when $N$ is very large, the cumulants of order $>2$ become negligible. Therefore, the distribution
of the sum is only characterised by its first two cumulants (mean and variance): it is a Gaussian.

Let us now turn to the case where the elementary distribution $P_1(x_1)$ decreases as a power
law for large arguments $x_1$  (cf.\ (\ref{E_ASYM})), with a certain exponent $\mu$.
The cumulants of order higher than  $\mu$ are thus divergent. By studying the
small $z$ singular expansion of the Fourier transform of $P(x,N)$, one finds
that the above additivity property of cumulants
is bequeathed to the tail amplitudes $A_\pm^\mu$: the asymptotic behaviour of the distribution
of the sum  $P(x,N)$ {\it still behaves as a power-law} (which is thus conserved by addition
for all values of $\mu$, provided one takes the limit  $x \to
\infty$ {\it before} $N \to \infty$ -- see the discussion in  \ref{S_GD}), 
with a tail amplitude given by:
\be
A_{\pm,N}^\mu \equiv N A_{\pm}^\mu.
\ee
The tail parameter thus play the role, for power-law variables, of a generalised 
cumulant.

\subsection{Stable distributions and self-similarity}
\label{S_LOISTABLE}
\label{loistable}
\begin{figure}
\centerline{\hbox{
\epsfig{figure=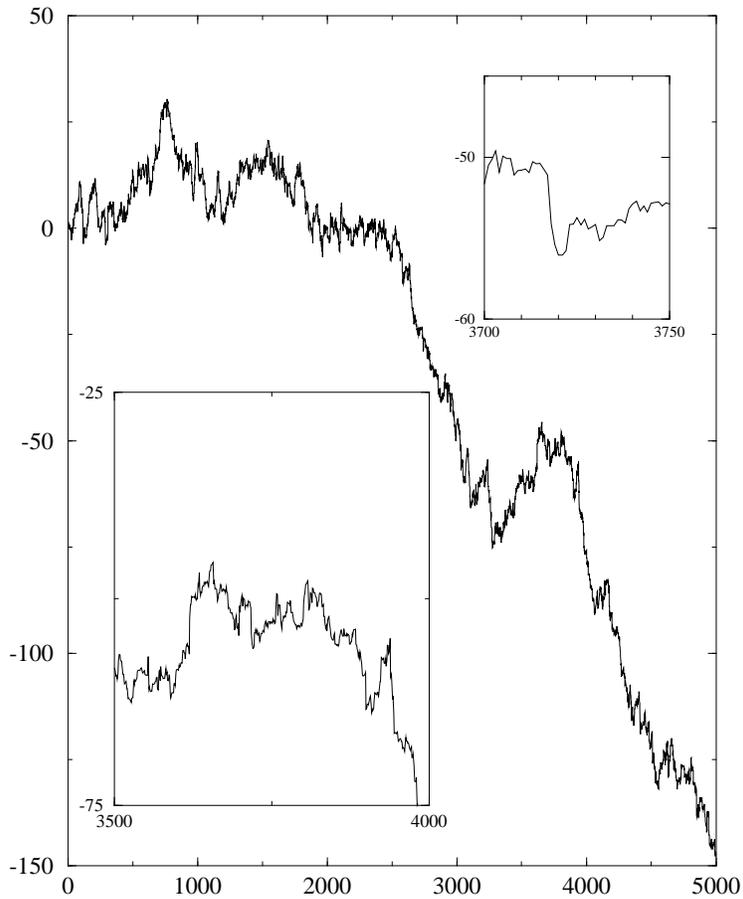,width=10cm}
}}
\vspace{0.5cm}\caption{Example of a self-affine function, obtained by summing random variables.
One plots the sum $x$ as a function of the number of terms $N$ in the
sum, for a Gaussian elementary distribution $P_1(x_1)$. Several
successive `zooms' reveal the self similar nature of the function,
here with $a_N = N^{1/2}$.}
\label{FigI5a}
\end{figure}

\begin{figure}
\centerline{\hbox{
\epsfig{figure=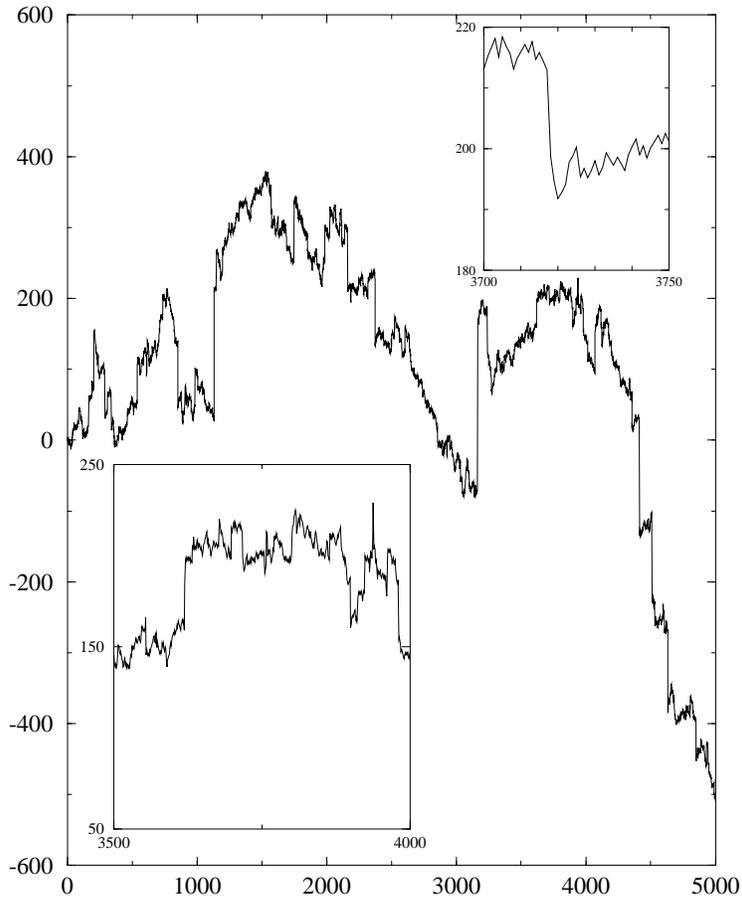,width=10cm}}}
\caption{In this case, the elementary distribution 
$P_1(x_1)$ decreases as a power-law with an exponent 
$\mu=1.5$. The scale factor is now given by $a_N=N^{2/3}$. Note that, contrarily to the previous
graph, one clearly observes the presence of sudden `jumps', which reflect the existence of very large
values of the elementary increment $x_1$.
 }
\label{FigI5b}
\end{figure}

If one adds random variables distributed according to an arbitrary law $P_1(x_1)$,
one constructs a random variable which has, in general, a different probability
distribution ($P(x,N)=[P_1(x_1)]^{\star N}$). However, for certain special 
distributions, the law of the sum has exactly the same shape as the elementary distribution --
these are called {\it stable} laws. The fact that two distributions have the `same shape'
means that one can find a ($N$ dependent) translation and dilation of $x$ such that the two laws coincide:
\be
P(x,N)dx = P_1(x_1)dx_1 \ \mbox{ where } x=a_N x_1+b_N .
\ee
The distribution of increments on a certain time scale (week, month,
year) is thus {\it scale invariant}, provided the variable $X$ is
properly rescaled. In this case, the chart giving the evolution of the
price of a financial asset as a function of time has the same
statistical structure, independently of the chosen elementary time
scale -- only the average slope and the amplitude of the fluctuations
are different. These charts are then called {\it self-similar}, or,
using a better terminology introduced by Mandelbrot, {\it self-affine}
(Figs.\ \ref{FigI5a} and
\ref{FigI5b}). \label{invariancedechelle}

The family of all possible stable laws coincide (for continuous
variables) with the L\'evy distributions defined above,\footnote{For
discrete variables, one should also add the Poisson distribution
(\ref{poisson}).}  which include Gaussians as the special case
$\mu=2$. This is easily seen in Fourier space, using the explicit
shape of the characteristic function of the L\'evy distributions. We
shall specialise here for simplicity to the case of symmetric
distributions $P_1(x_1)=P_1(-x_1)$, for which the translation factor
is zero ($b_N \equiv 0$). The scale parameter is then given by $a_N =
N^{\frac{1}{\mu}}$,\footnote{The case $\mu=1$ is special and involves
extra logarithmic factors.}  and one finds, for $\mu < 2$:
\be\label{propto}
\langle |x|^q \rangle ^{\frac{1}{q}} \propto A N^{\frac{1}{\mu}} 
\qquad q < \mu 
\ee
where $A=A_+=A_-$.  In words, the above equation means that the order
of magnitude of the fluctuations on `time' scale $N$ is a factor
$N^{\frac{1}{\mu}}$ larger than the fluctuations on the elementary
time scale. However, once this factor is taken into account, the
probability distributions are identical. One should notice the smaller
the value of $\mu$, the faster the growth of fluctuations with time.

\section{Central limit theorem}
\label{theoremelimitecentrale}

We have thus seen that the stable laws (Gaussian and L\'evy distributions) are `fixed
points' of the convolution operation. These fixed points are actually also {\it attractors},
in the sense that any distribution convoluted with itself a large number of times finally
converges towards a stable law (apart from some very pathological cases). Said differently,
the limit distribution of the sum of a large number of random variables is a stable law.
The precise formulation of this result is known as the {\it central limit theorem} ({\sc clt}).

\subsection{Convergence to a Gaussian}

The classical formulation of the {\sc clt} deals with sums of {\sc iid} random variables 
of {\it finite variance $\sigma^2$} towards a Gaussian. In a more precise way, the result is then 
the following:
 \be
\lim_{N\to \infty} {\cal P}\left(u_1 \leq \frac{x-m N}{\sigma \sqrt{N}} \leq
u_2\right)= \int_{u_1}^{u_2} \frac{du}{\sqrt{2\pi}} e^{-u^2/2},
\ee
for all {\it finite} $u_1,u_2$. Note however that for finite $N$, the distribution of
the sum $X=X_1+...+X_N$ in the tails (corresponding to extreme events) can be very different from the
Gaussian prediction; but the weight of these non-Gaussian regions tends to zero when
$N$ goes to infinity. The {\sc clt} only concerns the {\it central} region, which keeps a finite 
weight for $N$ large: we shall come back in detail to this point below.

The main hypotheses insuring the validity of the Gaussian {\sc clt} are the following:

\begin{itemize}

\item The $X_i$ must be independent random variables, or at least not `too' correlated (the
correlation function 
$\langle x_i x_j \rangle - m^2$ must decay sufficiently fast when $|i-j|$ becomes large -- see
1.7.1 below).  For example, in the extreme case where all the $X_i$ are perfectly correlated (i.e.\ 
they are all equal), the distribution of $X$ is obviously the same as that of the individual $X_i$
(once the factor $N$ has been properly taken into account).

\item The random variables $X_i$ need not necessarily be identically distributed. One must
however require that the variance of all these distributions are not
too dissimilar, so that no one of the variances dominates over all the
others (as would be the case, for example, if the variances were
themselves distributed as a power-law with an exponent $\mu < 1$). In
this case, the variance of the Gaussian limit distribution is the
average of the individual variances. This also allows one to deal with
sums of the type $X=p_1 X_1+ p_2 X_2 + ...+p_N X_N$, where the $p_i$
are arbitrary coefficients; this case is relevant in many
circumstances, in particular in the Portfolio theory (cf.\ Chapter 3).

\item Formally, the {\sc clt} only applies in the limit where $N$ is infinite. In practice, $N$ 
must be large enough for a Gaussian to be a good approximation of the
distribution of the sum.  The minimum required value of $N$ (called
$N^*$ below) depends on the elementary distribution $P_1(x_1)$ and its
distance from a Gaussian. Also, $N^*$ depends on how far in the tails
one requires a Gaussian to be a good approximation, which takes us
to the next point.

\item As mentioned above, the {\sc clt} does not tell us anything about the tails of
the distribution of $X$; only the central part of the distribution is well described
by a Gaussian. The `central' region means a region of width at least on the order of
 $\sqrt{N}\sigma$ around the mean value of $X$. The actual width of the region where the 
Gaussian turns out to be a good approximation for large finite $N$ crucially depends on the
elementary distribution $P_1(x_1)$. This problem will be explored in Section \ref{S_GD}.
Roughly speaking, this region is of width 
$\sim N^{3/4}\sigma$ for `narrow' symmetric elementary distributions, such that all
even moments are finite. This region is however sometimes of much smaller extension: for
example, if  $P_1(x_1)$ has power-law tails with  $\mu > 2$ (such that $\sigma$ is
finite), the Gaussian `realm'  grows barely faster than $\sqrt{N}$ (as
 $\sim \sqrt{N\log N}$).

\end{itemize}

\begin{technical}
The above formulation of the {\sc clt} requires the existence of a
finite variance.  This condition can be somewhat weakened to include
some `marginal' distributions such as a power-law with $\mu=2$. In
this case the scale factor is not $a_N=\sqrt{N}$ but rather
$a_N=\sqrt{N \ln N}$. However, as we shall discuss in the next
section, elementary distributions which decay more slowly than
$|x|^{-3}$ do not belong the the Gaussian basin of attraction. More
precisely, the necessary and sufficient condition for $P_1(x_1)$ to
belong to this basin is that:
\be 
\lim_{u\rightarrow\infty}u^2 \frac{{\cal P}_{1<}(-u)+{\cal P}_{1>}(u)} 
{\int_{|u'|<u}du'\ u'^2P_1(u')}=0.
\ee
This condition is always satisfied if the variance is finite, but allows one to
include the marginal cases such as a power-law with $\mu=2$.
\end{technical}

\subsubsection{The central limit theorem and information theory}
\begin{technical}
\label{information}
It is interesting to notice that the Gaussian is the law of maximum entropy -- or
minimum information -- such that its variance is fixed. The missing information quantity
 $\cal I$ (or entropy) associated with a probability distribution 
$P$ is defined as: 
\be
{\cal I}[P] \equiv - \int dx \ P(x) \ \log\left[\frac{P(x)}{e}\right].
\ee
The distribution maximising  ${\cal I}[P]$ for a given value of the variance is obtained 
by taking a functional derivative with respect to $P(x)$:
\be
\frac{\partial}{\partial P(x)} \left [{\cal I}[P] - \zeta \int dx' \ 
x'^2 P(x') - \zeta' \int dx' \ P(x') \right] = 0,
\label{E_ENTROPIE} \ee
where  $\zeta$ is fixed by the condition $ \int dx \ 
x^2 P(x) = \sigma^2$ and $\zeta'$ by the normalisation of $P(x)$. It is immediate to show
that the solution to  (\ref{E_ENTROPIE}) is indeed the Gaussian. The numerical value of its entropy 
is:
\be
{\cal I}_G = \frac{3}{2} + \frac{1}{2} \log(2\pi) + \log(\sigma) \simeq 2.419 +
\log(\sigma).
 \ee
For comparison, one can compute the entropy of the symmetric exponential distribution, which is:
\be
{\cal I}_E = 2 + \frac{\log 2}{2} + \log(\sigma) \simeq 2.346 + \log(\sigma).
\ee
It is important to understand that the convolution operation is `information burning',
since all the details of the elementary distribution $P_1(x_1)$ progressively disappear
while the Gaussian distribution emerges.
\end{technical}

\subsection{Convergence to a L\'evy distribution}
\label{S_TLCLEVY}

Let us now turn to the case of the sum of a large number $N$ of {\sc iid} random
variables, asymptotically distributed as a power-law with $\mu<2$, and
with a tail amplitude  $A^\mu=A^\mu_{+}=A^\mu_{-}$ (cf.\ (\ref{E_ASYM})). The variance
of the distribution is thus infinite. The limit distribution for large $N$ is then
a stable L\'evy distribution of exponent $\mu$ and with a tail amplitude $NA^\mu$.  
If the positive and negative tails of the elementary distribution $P_1(x_1)$
are characterised by different amplitudes 
($A^\mu_-$ and $A^\mu_+$) one then obtains an asymmetric L\'evy distribution with parameter
$\beta=(A^\mu_+-A^\mu_-)/(A^\mu_++A^\mu_-)$. If the `left' exponent is different from the `right'
exponent ($\mu_- \neq \mu_+$), then the smallest of the two wins and one finally obtains 
a totally asymmetric L\'evy distribution  ($\beta=-1$ or  $\beta=1$) with exponent 
$\mu=\min(\mu_-,\mu_+)$. The {\sc clt} generalised to L\'evy distributions applies with the same precautions as in the Gaussian case above.

\begin{technical}
Technically, a distribution  $P_1(x_1)$ belongs to the attraction basin 
of the L\'evy distribution  $L_{\mu,\beta}$
if and only if:
\be
\lim_{u\rightarrow\infty}
\frac{{\cal P}_{1<} (-u)}{{\cal P}_{1>} (u)}=\frac{1-\beta}{1+\beta}  ;
\ee
and for all $r$,
\be
\lim_{u\rightarrow\infty}
\frac{{\cal P}_{1<} (-u)+{\cal P}_{1>} (u)}{{\cal P}_{1<} (-ru)+{\cal P}_{1>} (ru)}=
r^{\mu}.
\ee
A distribution with an asymptotic tail given by (\ref{E_ASYM}) is such that,
\be
{\cal P}_{1<}(u)\mathop\simeq_{u\rightarrow -\infty} \frac{A^\mu_{-}}{|u|^{\mu}}
\mbox{ and }
{\cal P}_{1>}(u)\mathop\simeq_{u\rightarrow \infty} \frac{A^\mu_{+}}{u^{\mu}},
\ee
and thus belongs to the attraction basin of the L\'evy distribution of exponent  $\mu$
and asymmetry parameter $\beta=(A^\mu_{+}-A^\mu_{-})/(A^\mu_{+}+A^\mu_{-})$.

\end{technical}

\subsection{Large deviations}
\label{S_GD}
\label{grandesdeviations}
The {\sc clt} teaches us that the Gaussian approximation is justified 
to describe the `central' part of the distribution of the sum of a large number 
of random variables (of finite variance). However, the definition of the {\it centre} has
remained rather vague up to now. The {\sc clt} only states that the probability of
finding an event in the {\it tails } goes to zero for large $N$.
In the present section, we characterise more precisely the region where the 
Gaussian approximation is valid. 

If $X$ is the sum of $N$  {\sc iid} random variables of mean $m$ and variance $\sigma^2$,
one defines a `rescaled variable' $U$ as:
\be
U=\frac{X-Nm}{\sigma \sqrt{N}},
\ee
which according to the  {\sc clt} tends towards a Gaussian variable 
of zero mean and unit variance. Hence, for any {\it fixed} \/ $u$, one has:
\be 
\lim_{N\rightarrow\infty}
{\cal P}_{>}(u)={\cal P}_{G>}(u),
\ee
where ${\cal P}_{G>}(u)$ is the related to the error function, and describes the
weight contained in the tails of the Gaussian:
\be
{\cal P}_{G>}(u)=\int^{\infty}_{u}\frac{du'}{\sqrt{2\pi}}\exp(-u^2/2)
=\tfrac{1}{2}\mbox{erfc}\left(\frac{u}{\sqrt{2}}\right).\label{erfc}
\ee
However, the above convergence is {\it not uniform}. The value of $N$ such that
the approximation ${\cal P}_{>}(u)\simeq {\cal P}_{G>}(u)$ becomes valid depends on $u$.
Conversely, for fixed $N$, this approximation is only valid for $u$ not too large: $|u|\ll u_0(N)$.

One can estimate $u_0(N)$ in the case where the elementary distribution 
$P_1(x_1)$ is `narrow', that is, decreasing faster than any power-law when  $|x_1|\rightarrow\infty$,
such that all the moments are finite. In this case, all the cumulants of $P_1$ are finite
and one can obtain a systematic expansion in powers of $N^{-1/2}$ of the difference 
 $\Delta{\cal 
P}_{>}(u)\equiv{\cal P}_{>}(u)-{\cal P}_{G>}(u)$,
\be\label{E_GD}
\Delta{\cal P}_{>}(u)\simeq
\frac{\exp(-u^2/2)}{\sqrt{2\pi}}\left(
\frac{Q_1(u)}{N^{1/2}}+
\frac{Q_2(u)}{N}+\ldots+
\frac{Q_k(u)}{N^{k/2}}+\ldots\right),
\ee
where the  $Q_k(u)$ are polynomials functions which can be expressed in terms of the 
normalised cumulants
$\lambda_n$ (cf.\ (\ref{E_CUMNORM})) of the elementary distribution. 
More explicitely, the first two terms are given by:
 \be
Q_1(u)=\tfrac{1}{6}\lambda_3(u^2-1) ,
\ee
and
\be 
Q_2(u)=\tfrac{1}{72}\lambda_3^2 u^5+
\tfrac{1}{8}(\tfrac{1}{3} \lambda_4-\tfrac{10}{9} \lambda_3^2)u^3+
(\tfrac{5}{24} \lambda_3^2-\tfrac{1}{8} \lambda_4)u.
\ee

One recovers the fact that if all the cumulants of $P_1(x_1)$ of order larger than two are zero,
all the $Q_k$ are also identically zero and so is the difference between $P(x,N)$ and
the Gaussian.

For a general asymmetric elementary distribution $P_1$,  $\lambda_3$ is non zero. 
The leading term in the above expansion when $N$ is large is thus $Q_1(u)$. For the
Gaussian approximation to be meaningful, one must at least require that this term is small
in the central region where $u$ is of order one, which corresponds to  $x - mN \sim\sigma \sqrt{N}$. 
This thus imposes that $N\gg N^*=\lambda_3^2$. The Gaussian 
approximation remains valid whenever the relative error is small compared to $1$. 
For large $u$ (which will be justified for large $N$), the relative error is obtained by dividing Eq.\ (\ref{E_GD}) by
 ${\cal P}_{G>}(u)\simeq \exp(-u^2/2)/(u\sqrt{2\pi})$. One then obtains the
following condition:\footnote{The above arguments can actually be made fully rigourous, 
see [Feller].}
\be\label{E_GDC3}
\lambda_3 u^3 \ll N^{1/2} \mbox{ i.e. }  
|x-N m| \ll \sigma \sqrt{N} \left(\frac{N}{N^*}\right)^{1/6}.
\ee
This shows that the central region has an extension growing as $N^{\frac{2}{3}}$.

A symmetric elementary distribution is such that $\lambda_3\equiv 0$; it is then the kurtosis
 $\kappa=\lambda_4$ that fixes the first correction to the Gaussian when $N$ is large, and
thus the extension of the central region. The conditions now read:
 $N \gg N^* =\lambda_4 $ and 
\be \label{E_GDC4} \lambda_4 u^4 \ll N
\mbox{ i.e. }   |x-N m| \ll \sigma \sqrt{N} \left(\frac{N}{N^*}\right)^{{1}/{4}} .
\ee
The central region now extends over a region of width 
$N^{3/4}$.

The results of the present section do not directly apply if the elementary distribution
 $P_1(x_1)$ decreases as a power-law (`broad distribution'). In this case, some of the cumulants
are infinite and the above cumulant expansion  (\ref{E_GD}) is meaningless. 
In the next section, we shall see that in this case the `central' region is much more
restricted than in the case of `narrow' distributions. We shall then describe in Section
\ref{S_LEVY_TRONQUEE}, the case of `truncated' power-law distributions, where the above
conditions become asymptotically relevant. These laws however may have a very large
kurtosis, which depends on the point where the truncation becomes noticeable, and the
above condition $N\gg \lambda_4$ can be hard to satisfy.

\subsubsection{Cram\`er function}\label{cramer}
\begin{technical}
More generally, when $N$ is large, one can write the distribution of the sum of 
$N$ {\sc iid} random variables as:\footnote{We assume that their mean is zero, which can always be
achieved through a suitable shift of $x_1$.} 
\be
P(x,N) \mathop{\simeq}_{N \to \infty}
\exp\left[-N S\left(\frac{x}{N}\right)\right],
\ee
where  $S$ is the so-called Cram\`er function, which gives some information about the 
probability of $X$ even outside the `central' region. When the variance is finite, $S$ 
grows as $S(u) \propto u^2$ for small $u$'s, which again leads to a Gaussian central region.
For finite $u$, $S$ can be computed using Laplace's saddle point method, valid for $N$ large.
By definition:  \label{col1}
\be\label{methode_col}
P(x,N)= \int \frac{dz}{2\pi} \exp N\left(-i z \frac{x}{N} + 
\log[\hat P_1(z)]\right).
\ee
When $N$ is large, the above integral is dominated by the neighbourhood of the point $z^*$
where the term in the exponential is stationary. The results can be written as:
\be
P(x,N)\simeq \exp\left[-N S\left(\frac{x}{N}\right)\right],\label{E_CRAMER}
\ee
with $S(u)$ given by:
\be
\left. \frac{d \log[\hat P_1(z)]}{dz}\right|_{z=z^*} = i u \qquad S(u)= -i z^* u + \log[\hat
P_1(z^*)],
\ee
which, in principle, allows one to estimate $P(x,N)$ even outside the central region. Note that 
if $S(u)$ is finite for finite $u$, the corresponding probability is exponentially small in $N$.

\end{technical}
\subsection{The CLT  at work on a simple case}

It is helpful to give some flesh to the above general statements, by working out
explicitly the convergence towards the Gaussian in two exactly soluble cases. On these examples,
one clearly sees the domain of validity of the {\sc clt} as well as its limitations.

Let us first study the case of positive random variables distributed according to the exponential
distribution:
\be
P_1(x)=\Theta(x_1)\alpha e^{-\alpha x_1}, 
\ee
where $\Theta(x_1)$ is the function equal to $1$ for $x_1\geq 0$ and to  0
otherwise. A simple computation shows that the above distribution is correctly 
normalised, has a mean given by  $m=\alpha^{-1}$ and a variance given by $\sigma^2=\alpha^{-2}$. 
Furthermore, the exponential distribution is asymmetrical; its skewness is given by
$c_3=\langle (x-m)^3 \rangle = 2 \alpha^{-3}$, or $\lambda_3=2$.

The sum of $N$ such variables is distributed according to the $N^{th}$ convolution
of the exponential distribution. According to the {\sc clt} this distribution 
should approach a Gaussian of mean $mN$ and of variance 
$N\sigma^2$. The $N^{th}$ convolution
of the exponential distribution can be computed exactly. The result is:\footnote{
This result can be shown by induction using the definition (\ref{E_CONV}).} 
\be\label{E_GAMMAN}
P(x,N)=\Theta(x)\alpha^N \frac{x^{N-1}e^{-\alpha x}}{(N-1)!} ,
\ee
which is called a `Gamma' distribution of index $N$. At first sight, this distribution 
does not look very much like a Gaussian! For example, its asymptotic behaviour is very far
from that of a Gaussian: the `left' side is strictly zero for negative $x$, while the
`right' tail is exponential, and thus much fatter than the Gaussian. It is thus
very clear that the {\sc clt} does not apply for values of $x$ too far from the mean value.
However, the central region around $Nm=N\alpha^{-1}$ is well described by a Gaussian. The
most probable value ($x^*$) is defined as:
\be
\left.\frac{d}{dx} x^{N-1}e^{-\alpha x} \right|_{x^*}=0,
\ee
or $x^*=(N-1)m$. An expansion in  $x-x^*$ of $P(x,N)$ then gives us:
\ba\nonumber
\log P(x,N)&=&-K(N-1)-\log m
-\frac{\alpha^2 (x-x^*)^2}{2(N-1)}\\ 
&&\mbox{}+\frac{\alpha^3 (x-x^*)^3}{3(N-1)^2}
+O(x-x^*)^4, 
\ea
where
\be
K(N) \equiv \log N! +N-N\log N\mathop{\simeq}_{N\rightarrow\infty}\tfrac{1}{2}
\log(2\pi N).
\ee
Hence, to second order in $x-x^*$, $P(x,N)$ is given by a Gaussian of mean  $(N-1)m$ and
 variance $(N-1)\sigma^2$. The relative difference between $N$ and  $N-1$ goes to zero
for large $N$. Hence, for the Gaussian approximation to be valid, one requires not only that
$N$ be large compared to one, but also that the higher order terms in  $(x-x^*)$ be negligible.
The cubic correction is small compared to $1$ as long as  $\alpha |x-x^*| \ll  N^{2/3}$, 
in agreement with the above general statement (\ref{E_GDC3}) for an elementary distribution 
with a non zero third cumulant. Note also that for $x \to \infty$, the 
exponential behaviour of the Gamma function coincides (up to subleading terms in $x^{N-1}$) with
the asymptotic behaviour of the elementary distribution $P_1(x_1)$.

Another very instructive example is provided by a distribution which behaves as a power-law
for large arguments, but at the same time has a finite variance to ensure the validity of
the {\sc clt}. Consider the following explicit example of a Student distribution
with $\mu=3$:\label{student2}
\be\label{E_LOIMU3}
P_1(x_1)=\frac{2 a^{3}}{\pi(x_1^2+a^{2})^{2}},
\ee
where $a$ is a positive constant. This symmetric distribution behaves as a power-law with $\mu=3$  (cf.\ (\ref{E_ASYM})); all its cumulants of order
larger or equal to three are infinite. However, its variance is finite and equal to $a^{2}$.
\begin{technical}
It is useful to compute the characteristic function of this distribution,
\be
\hat{P_1}(z)=(1+a|z|)e^{-a|z|},
\ee
and the first terms of its small $z$ expansion, which read:
\be
\hat{P_1}(z)\simeq 1-\frac{z^{2}a^{2}}{2}+\frac{|z|^{3}a^{3}}{3}+O(z^{4}).
\ee
The first singular term in this expansion is thus  $|z|^{3}$, as expected
from the asymptotic behaviour of  $P_1(x_1)$ in  $x_1^{-4}$, and the
divergence of the moments of order larger than three.

The $N^{th}$ convolution of $P_1(x_1)$ thus has the following characteristic 
function:
\be
\hat{P_1}^{N}(z)=(1+a|z|)^{N}e^{-aN|z|},
\ee
which, expanded around $z=0$, gives:
\be
\hat{P_1}^{N}(k)\simeq 1-\frac{N z^{2}a^{2}}{2}+\frac{N |z|^{3}a^{3}}{3}+O(z^{4}).
\ee
Note that the  $|z|^3$ singularity  (which signals the divergence of the moments  $m_n$ for $n\geq 3$)
{\it does not disappear} under convolution, even if at the same time $P(x,N)$ converges
towards the Gaussian. The resolution of this apparent paradox is again that the 
convergence towards the Gaussian only concerns the centre of the distribution, while the
tail in  $x^{-4}$ survives for ever (as was mentioned in Section \ref{S_LOISTABLE}). 
\end{technical}

As follows from the {\sc clt},  the centre of  $P(x,N)$ is well approximated, for $N$ large,
by a Gaussian of zero mean and variance  
$N a^{2}$:
\be\label{E_MU3GAUSS}
P(x,N)\simeq
\frac{1}{\sqrt{2\pi N}a}
\exp\left(-\frac{x^{2}}{2Na^{2}}\right).
\ee
On the other hand, since the power-law behaviour is conserved upon addition and that
the tail amplitudes simply add (cf.\ (\ref{E_ASYM})), one also has, for large $x$'s:
\be\label{E_MU3ASYM}
P(x,N)\mathop{\simeq}_{x\rightarrow\infty} \frac{2Na^{3}}{\pi x^{4}}.
\ee
The above two expressions (\ref{E_MU3GAUSS}) and (\ref{E_MU3ASYM}) are
not incompatible, since these describe two very different regions of
the distribution $P(x,N)$. For fixed $N$, there is a characteristic
value $x_{0}(N)$ beyond which the Gaussian approximation for $P(x,N)$
is no longer accurate, and the distribution is described by its
asymptotic power-law regime.  The order of magnitude of $x_{0}(N)$ is
fixed by looking at the point where the two regimes match to one
another:
\be
\frac{1}{\sqrt{2\pi N}a}
\exp\left(-\frac{x_{0}^{2}}{2Na^{2}}\right)\simeq
\frac{2Na^{3}}{\pi x_{0}^{4}}.
\ee
One thus find,
\be
x_{0}(N)\simeq a\sqrt{N\log N},
\ee
(neglecting subleading corrections for large $N$). 

This means that the rescaled variable $U=X/(a\sqrt{N})$ becomes for
large $N$ a Gaussian variable of unit variance, but this description
ceases to be valid as soon as $u \sim \sqrt{\log N}$, which grows very
slowly with $N$. For example, for $N$ equal to a million, the Gaussian
approximation is only acceptable for fluctuations of $u$ of less than
three or four {\sc rms}!

Finally, the {\sc clt} states that the weight of the regions where
$P(x,N)$ substantially differs from the Gaussian goes to zero when $N$
becomes large. For our example, one finds that the probability that $X$
falls in the tail region rather than in the central region is given
by:
\be
{\cal P}_<(x_0)+{\cal P}_>(x_0)\simeq 2\int_{a\sqrt{N\log N}}^{\infty}
\frac{2 a^{3}N}{\pi x^{4}} dx \propto \frac{1}{\sqrt{N}\log^{3/2}N},
\ee
which indeed goes to zero for large $N$.

The above arguments are not special to the case $\mu=3$ and in fact apply more generally,
as long as $\mu > 2$, i.e.\ when the variance is finite. In the general case, one finds
that the {\sc clt} is valid in the region $|x|\ll x_0\propto\sqrt{N\log N}$, and that the
weight of the non Gaussian tails is given by:
\be
{\cal P}_<(x_0)+{\cal P}_>(x_0)\propto  
\frac{1}{ N^{\mu/2-1}\log^{\mu/2}N},
\ee
which tends to zero for large $N$. However, one should notice that as $\mu$ approaches
the `dangerous' value $\mu=2$, the weight of the tails becomes more and more important. 
For $\mu < 2$, the whole argument collapses since the weight of the tails would grow with $N$.
In this case, however, the convergence is no longer towards the Gaussian, but towards the
L\'evy distribution of exponent $\mu$.

\subsection{Truncated L\'evy distributions}
\label{S_LEVY_TRONQUEE}
\label{loidelevytronquee2}
An interesting case is when the elementary distribution $P_1(x_1)$ is a truncated 
L\'evy distribution ({\sc tld}) as defined in Section 1.3.3. The first
cumulants of the distribution defined by Eq.\ (\ref{E_FLEVYT}) read, for  $1 < \mu < 2$:
\be\label{E_SIGLT}
c_2 = \mu (\mu-1) \frac{a_\mu}{|\cos{\pi \mu/2}|} \alpha^{\mu-2} \qquad c_3=0. 
\ee
The kurtosis $\kappa=\lambda_4 = {c_4}/{c_2^2}$ is given by:
 \be\label{E_KURTOLT}
\lambda_4 = \frac{(3-\mu)(2-\mu)|\cos{\pi \mu/2}|}{\mu (\mu-1) a_\mu \alpha^\mu}.
\ee
Note that the case $\mu=2$ corresponds to the Gaussian, for which $\lambda_4=0$ as expected. 
On the other hand, when  $\alpha \to 0$, one recovers a pure L\'evy distribution, for which
$c_2$ and $c_4$ are formally infinite. Finally, if $\alpha \to \infty$ with
$a_\mu \alpha^{\mu-2}$ fixed, one also recovers the Gaussian.

\begin{figure}
\centerline{\hbox{
\epsfig{figure=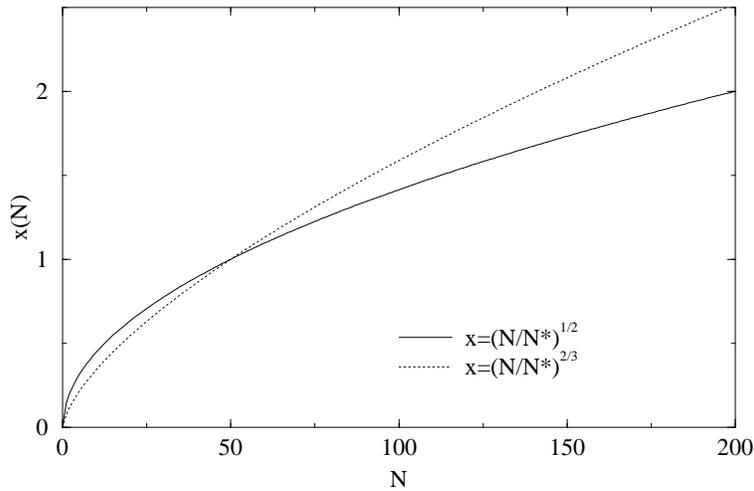,width=10cm}
}}
\caption{Behaviour of the typical value of $X$ as a function of $N$ for {\sc tld} variables.
When  $N \ll N^*$, $x$ grows  $N^{\frac{1}{\mu}}$ 
(dotted line).  When $N \sim N^*$, $x$ reaches the value 
$\alpha^{-1}$ and the exponential cut-off starts being relevant. 
When $N \gg N^*$, the behaviour predicted by the {\sc clt} sets in, and one recovers 
 $x \propto \protect\sqrt{N}$ (plain line).  }
\label{FigI6}
\end{figure}

If one considers the sum of $N$ random variables distributed according to a {\sc tld}, the 
condition for the {\sc clt} to be valid reads (for $\mu < 2$):\footnote{One can see by inspection that the
other conditions, concerning higher order cumulants, and which read
$N^{k-1} \lambda_{2k} \gg 1$, are actually equivalent to the one written here.}   
 \be  
N \gg N^*=\lambda_4 \Longrightarrow (N a_\mu)^{\frac{1}{\mu}} \gg \alpha^{-1}.
\ee
This condition has a very simple intuitive meaning. A {\sc tld}
behaves very much like a pure L\'evy distribution as long as $x \ll
\alpha^{-1}$.  In particular, it behaves as a power-law of exponent
$\mu$ and tail amplitude $A^\mu \propto a_\mu$ in the region where $x$
is large but still much smaller than $\alpha^{-1}$ (we thus also
assume that $\alpha$ is very small).  If $N$ is not too large, most
values of $x$ fall in the L\'evy-like region. The largest value of $x$
encountered is thus of order $x_{\max} \simeq A N^{\frac{1}{\mu}}$
(cf. \ref{E_MAXPOWER}).  If $x_{\max}$ is very small compared to
$\alpha^{-1}$, it is consistent to forget the exponential cut-off and
think of the elementary distribution as a pure L\'evy
distribution. One thus observe a first regime in $N$ where the typical
value of $X$ grows as $N^{\frac{1}{\mu}}$, as if $\alpha$ was zero.\footnote{Note however that
the variance of $X$ grows like $N$ for all $N$. However, the variance
is dominated by the cut-off and, in the region $N \ll N^*$, grossly
overestimates the typical values of $X$ -- see Section 2.3.2.} 
However, as illustrated in Fig.\ \ref{FigI6}, this regime ends when
$x_{\max}$ reaches the cut-off value $\alpha^{-1}$: this happens
precisely when $N$ is on the order of $N^*$ defined above. For $N >
N^*$, the variable $X$ progressively converges towards a Gaussian
variable of width $\sqrt{N}$, at least in the region where $|x|\ll \sigma
N^{3/4}/N^{*1/4}$.  The typical amplitude of $X$ thus behaves (as a
function of $N$) as sketched in Fig.\ \ref{FigI6}. Notice that the
asymptotic part of the distribution of $X$ (outside the central
region) decays as an exponential for all values of $N$.

\subsection{Conclusion: survival and vanishing of tails}
\label{queue2}
The {\sc clt} thus teaches us that if the number of terms in a sum is large, the
sum becomes (nearly) a Gaussian variable. This sum can represent the temporal  aggregation
of the daily fluctuations of a financial asset, or the aggregation, in a portfolio, of different
stocks. The Gaussian (or non-Gaussian) nature of this sum is thus of crucial importance for
risk control, since the extreme tails of the distribution correspond to the most `dangerous'
fluctuations. As we have discussed above, fluctuations are never Gaussian in the far-tails:
one can explicitly show that if the elementary distribution decays as a power-law (or as
an exponential, which formally corresponds to $\mu=\infty$), the distribution of the sum
decays in the very same manner outside the central region, i.e.\ much more slowly than 
the Gaussian. The {\sc clt} simply ensures that these tail regions are expelled more and more
towards large values of $X$ when $N$ grows, and their associated probability is smaller and smaller.
When confronted to a concrete problem, one must decide whether $N$ is large enough to be
satisfied with a Gaussian description of the risks. In particular, if $N$ is less than the
characteristic value $N^*$ defined above, the Gaussian approximation is very bad.

\section{Correlations, dependence and non-stationary models (*)}
\label{S_NONSTAT}
\label{nonstat}
We have assumed up to now that the random variables where {\it independent} and {\it identically
distributed}. Although the general case cannot be discussed as thoroughly as the {\sc iid} case,
it is useful to illustrate how the {\sc clt} must be modified on a few examples, some of which
being particularly relevant in the context of financial time series.\label{iid2}

\subsection{Correlations}

Let us assume that the correlation function $C_{i,j}$ (defined as
$\langle x_i x_j \rangle-m^2$) of the random variables is non zero for
$i \neq j$. We also assume that the process is {\it stationary},
i.e.\ that $C_{i,j}$ only depends on $|i-j|$: $C_{i,j}=C(|i-j|)$, with
$C(\infty)=0$. The variance of the sum can be expressed in terms of
the matrix $C$ as:\footnote{We again assume in the following, without
loss of generality that the mean $m$ is zero.}
\be
\langle x^2 \rangle = \sum_{i,j=1}^N C_{i,j} = N \sigma^2 + 2N \sum_{\ell=1}^N (1-\frac{\ell}{N}) C(\ell)
\ee
where $\sigma^2 \equiv C(0)$.
From this expression, it is readily seen that if $C(\ell)$ decays faster than $1/\ell$ for large $\ell$,
the sum over $\ell$ tends to a constant for large $N$, and thus the variance of the sum still grows 
as $N$, as for the usual {\sc clt}. If however $C(\ell)$ decays for large $\ell$ as a power-law 
$\ell^{-\nu}$, with $\nu < 1$, then the variance grows faster than $N$, as $N^{2-\nu}$  -- correlations
thus enhance fluctuations. Hence, when $\nu < 1$, the standard
{\sc clt} certainly has to be amended. The problem of the limit distribution in these cases is however
not solved in general. For example, if the $X_i$ are correlated Gaussian variables, it is easy to
show that the resulting sum is also Gaussian, whatever the value of $\nu$. Another solvable case is when
the $X_i$ are correlated Gaussian variables, but one takes the sum of the {\it squares} of the $X_i$'s. 
This sum converges towards a Gaussian of width $\sqrt{N}$ whenever $\nu > 1/2$, but towards a non trivial
limit distribution of a new kind (i.e.\ neither Gaussian nor L\'evy stable) when $\nu < 1/2$. In this last
case, the proper rescaling factor must be chosen as $N^{1-\nu}$.

One can also construct {\it anti-correlated} random variables, the sum of which  grows slower than $\sqrt{N}$.
In the case of power-law correlated or anticorrelated Gaussian random variables, one speaks of
`fractional Brownian motion'. This notion was introduced by Mandelbrot and
Van Ness [Mandelbrot].

\subsection{Non stationary models and dependence}

It may happen that the distribution of the elementary random variables
$P_1(x_1)$, $P_2(x_2), ...,P_N(x_N)$ are not all identical. This is the case, for
example, when the variance of the random process depends upon time -- in
financial markets, it is a well known fact that the daily volatility is time dependent,
taking rather high levels in periods of uncertainty, and reverting back to lower values
in calmer periods. For example, the volatility of the bond market has been very
high during 1994, and decreased in later years. Similarly, the volatility of
stock markets has increased since August 1997.

If the distribution $P_k$ varies sufficiently `slowly', one can in principle measure some of
its moments (for example its mean and variance) over a time scale which is long enough to allow
for a precise determination of these moments, but short compared to the time scale over which
$P_k$ is expected to vary. The situation is less clear if $P_k$ varies `rapidly'. Suppose for
example that 
$P_k(x_k)$ is a Gaussian distribution of variance $\sigma^2_k$, which is itself a random variable. We shall
denote as  $\overline{(...)}$ the average over the random variable $\sigma_k$, to distinguish it 
from the notation  $\langle 
...  \rangle_k$ which we have used to describe the average over the probability distribution $P_k$.
If $\sigma_k$ varies rapidly, it is impossible to separate the two sources of uncertainty. Thus, the
empirical histogram constructed from the series 
$\{x_1,x_2,...x_N\}$ leads to an `apparent' distribution $\overline 
P$ which is non-Gaussian even if each individual $P_k$ is Gaussian. Indeed, from:
\be
\overline{P}(x) \equiv \int d\sigma P(\sigma) \frac{1}{\sqrt{2 \pi \sigma^2}} \exp
-\frac{x^2}{2 \sigma^2},
\ee
one can calculate the kurtosis of $\overline P$ as:
\be
\overline{\kappa} = \frac{\overline{\langle x^4 \rangle}}{(\overline{\langle x^2
\rangle})^2} - 3 \equiv 3 \left(\frac{\overline{\sigma^4}}{(\overline{\sigma^2})^2}
-1 \right).
\ee
Since for any random variable one has $\overline{\sigma^4} \geq (\overline{\sigma^2})^2$ 
(the equality being reached only if $\sigma^2$ does not fluctuate at all), one finds that $\overline{\kappa}$
is always positive. The volatility fluctuations can thus lead to `fat tails'. More precisely, let us assume
that the
probability distribution of the {\sc rms}, 
$P(\sigma)$, decays itself for large $\sigma$ as $\exp -\sigma^c$, $c
>0$. Assuming $P_k$ to be Gaussian, it is easy to obtain, using a saddle point method
 (cf. (\ref{methode_col})), \label{col2} that for large $x$ one has:
\be
\log[\overline{P}(x)] \propto - x^{\frac{2c}{2+c}}.\label{strexp}
\ee
Since $c < 2+c$, this asymptotic decay is always much slower than in the Gaussian case, which corresponds to
$c \to \infty$. The case where the volatility itself has a Gaussian tail 
$(c=2)$ leads to an exponential decay of $\overline{P}(x)$.

Another interesting case is when $\sigma^2$ is distributed as an completely asymmetric L\'evy distribution
($\beta=1$) of exponent  $\mu < 1$. Using the properties of L\'evy distributions, one can then show that
$\overline{P}$ is itself a symmetric L\'evy distribution $(\beta=0)$,
of exponent equal to  $2 \mu$.

\label{volatstoch}

If the fluctuations of  $\sigma_k$ are themselves correlated, one observes
an interesting case of {\it dependence}. For example, if  $\sigma_k$ is
large, $\sigma_{k+1}$ will probably also be large. The fluctuation $X_k$ thus has
a large probability to be large (but of arbitrary sign) twice in a row. 
We shall often refer, in the following, to a simple model where $x_k$ can be
written as a product $\epsilon_k \sigma_k$, where $\epsilon_k$ are {\sc iid} random
variables of zero mean and unit variance, and $\sigma_k$ corresponds to the local `scale' of the fluctuations,
which can be correlated in time. The correlation function of the $X_k$ is thus given 
by:
\be
\overline{\langle x_i x_j \rangle} = \overline{\sigma_i \sigma_j}
\langle \epsilon_i \epsilon_j \rangle = \delta_{i,j}
\overline{\sigma^2}.\label{deltaK}
\ee
Hence the $X_k$ are uncorrelated random variables, but they are not independent since a higher
order correlation function reveals a richer structure. Let us for example consider the 
correlation of $X_k^2$:
\be
\overline{\langle x_i^2 x_j^2 \rangle} -
 \overline{\langle x_i^2\rangle} \overline{\langle x_j^2\rangle}=
 \overline{\sigma_i^2 \sigma_j^2}-
\overline{\sigma_i^2}\, \overline{\sigma_j^2} \qquad (i \neq j),
\ee
which indeed has an interesting temporal behaviour: see Section 2.4.\footnote{
Note that for $i\neq j$ this correlation function can be zero either because $\sigma$ is 
identically equal to a certain value $\sigma_0$, or because the fluctuations of $\sigma$ are
completely uncorrelated from one time to the next.} However, even if the correlation
function  $\overline{\sigma_i^2 \sigma_j^2}-\overline{\sigma}^2$  decreases very slowly 
with $|i-j|$, one can show that the sum of the  $X_k$, obtained as 
$\sum_{k=1}^N \epsilon_k \sigma_k$ is still governed by the {\sc clt}, and converges for
large $N$ towards a Gaussian variable. A way to see this is to compute the average kurtosis 
of the sum, $\kappa_N$. As shown in Appendix A, one finds the following result:
\be
\kappa_N = \frac{1}{N} \left[ \kappa_0 + 
(3+\kappa_0) g(0)+ 6 \sum_{\ell=1}^{N} (1-\frac{\ell}{N}) 
g(\ell)\right],
\ee
where $\kappa_0$ is the kurtosis of the variable $\epsilon$, and $g(\ell)$ the correlation
function of the variance, defined as:
\be 
\overline{\sigma_i^2 \sigma_j^2}-\overline{\sigma}^2=\overline{\sigma}^2 g(|i - j|)
\ee 
It is interesting to see that for $N=1$, the above formula gives 
$\kappa_1=\kappa_0+(3+\kappa_0) g(0) > \kappa_0$, which means that even if $\kappa_0 = 0$, 
a fluctuating volatility is enough to produce some kurtosis. More importantly, one
sees that if the variance correlation function $g(\ell)$ decays with $\ell$, the 
kurtosis $\kappa_N$ tends to zero with $N$, thus showing that the sum indeed converges towards 
a Gaussian variable. For example, if $g(\ell)$ decays as a power-law
$\ell^{-\nu}$ for large $\ell$, one finds that for large $N$:
\be
\kappa_N \propto \frac{1}{N} \quad {\rm for} \quad \nu > 1; \qquad
\kappa_N \propto \frac{1}{N^\nu} \quad {\rm for} \quad \nu < 1.
\ee
Hence, long-range correlation in the variance considerably slows down the convergence
towards the Gaussian. This remark will be of importance in the following, since
financial time series often reveal long-ranged volatility fluctuations.

\section{Central limit theorem for random matrices (*)}
\label{randommatrices}\label{eigen1}

One interesting application of the {\sc clt} concerns the spectral
properties of `random matrices'. The theory of Random Matrices has
made enormous progress during the past thirty years, with many
applications in physical sciences and elsewhere. More recently, it has
been suggested that random matrices might also play an important role
in finance: an example is discussed in Section 2.7. It is therefore
appropriate to give a cursory discussion of some salient properties of
random matrices. The simplest ensemble of random matrices is one where
a all elements of the matrix $\bf H$ are {\sc iid} random variables,
with the only constraint that the matrix be symmetrical
($H_{ij}=H_{ji}$). One interesting result is that in the limit of very
large matrices, the distribution of its eigenvalues has universal
properties, which are to a large extent independent of the
distribution of the elements of the matrix. This is actually the
consequence of the {\sc clt}, as we will show below. Let us introduce
first some notations.  The matrix $\bf H$ is a square, $N \times N$
symmetric matrix. Its eigenvalues are $\lambda_\alpha$, with
$\alpha=1,...,N$. The {\it density} of eigenvalues is defined as:
\be\label{eigenvalue}
\rho(\lambda)=\frac{1}{N} \sum_{\alpha=1}^N \delta(\lambda-\lambda_\alpha),
\ee
where $\delta$ is the Dirac function. We shall also need the so-called
`resolvent'\label{resolvent} ${\bf G}(\lambda)$ of the matrix $\bf H$,
defined as:
\be
G_{ij}(\lambda)\equiv\left(\frac{1}{\lambda {\bf 1} - {\bf H}}\right)_{ij},
\ee
where $\bf 1$ is the identity matrix. The trace of ${\bf G}(\lambda)$ can be expressed
using the eigenvalues of $\bf H$ as:
\be
\mbox{Tr} {\bf G}(\lambda)=\sum_{\alpha=1}^N \frac{1}{\lambda - \lambda_\alpha}.
\ee
The `trick' that allows one to calculate $\rho(\lambda)$ in the large $N$ limit is the 
following representation of the $\delta$ function:
\be
\frac{1}{x - i\epsilon} = PP \frac{1}{x} + i \pi \delta(x)  \qquad (\epsilon \to 0),
\ee
where $PP$ means the principal part. Therefore, $\rho(\lambda)$ can be expressed as:
\be
\rho(\lambda) = \lim_{\epsilon \to 0} \frac{1}{\pi} \Im\left(\mbox{Tr} {\bf G}(\lambda-i\epsilon)
\right).
\ee
Our task is therefore to obtain an expression for the resolvent ${\bf
G}(\lambda)$. This can be done by establishing a recursion relation,
allowing one to compute ${\bf G}(\lambda)$ for a matrix $\bf H$ with
one extra row and one extra column, the elements of which being
$H_{0i}$. One then computes $G_{00}^{N+1}(\lambda)$ (the superscript
stands for the size of the matrix $\bf H$) using the standard formula
for matrix inversion:
\be
G_{00}^{N+1}(\lambda) = \frac{{\mbox{minor}}(\lambda {\bf 1}- {\bf H})_{00}}{\det 
(\lambda {\bf 1}- {\bf H})}.
\ee
Now, one expands the determinant appearing in the denominator in minors along the first
row, and then each minor is itself expanded in subminors along their first column. After 
a little thought, this finally leads to the following expression for $G_{00}^{N+1}(\lambda)$:
\be
\frac{1}{G_{00}^{N+1}(\lambda)}=\lambda-H_{00} - \sum_{i,j=1}^{N} H_{0i}H_{0j} G_{ij}^N(\lambda). \label{recursionrelation}
\ee
This relation is general, without any assumption on the $H_{ij}$. Now, we assume that
the $H_{ij}$'s are {\sc iid} random variables, of zero mean and variance equal to $\langle H_{ij}^2 \rangle=\sigma^2/N$.
This scaling with $N$ can be understood as follows: when
the matrix $H$ acts on a certain vector, each component of the image vector is a sum of $N$
random variables. In order to keep the image vector (and thus the corresponding eigenvalue)
finite when $N \to \infty$, one should scale the elements of the matrix with the factor $1/\sqrt{N}$. 

One could also write a recursion relation for $G_{0i}^{N+1}$, and establish self-consistently
that $G_{ij} \sim 1/\sqrt{N}$ for $i \neq j$. On the other hand, due to the diagonal term
$\lambda$, $G_{ii}$ remains finite for $N \to \infty$. This scaling allows us to discard
all the terms with $i \neq j$ in the sum appearing in the right hand side of Eq.\ (\ref{recursionrelation}). Furthermore, since $H_{00} \sim 1/\sqrt{N}$, this term can
be neglected compared to $\lambda$. This finally leads to a simplified recursion relation,
valid in the limit $N \to \infty$:
\be
\frac{1}{G_{00}^{N+1}(\lambda)} \simeq \lambda - \sum_{i=1}^{N} H_{0i}^2 G_{ii}^N(\lambda). \label{recursionrelation2}
\ee
Now, using the {\sc clt}, we know that the last sum converges, for
large $N$, towards $\sigma^2/N \ \sum_{i=1}^{N}
G_{ii}^N(\lambda)$. This result is independent of the precise
statistics of the $H_{0i}$, provided their variance is
finite.\footnote{The case of L\'evy distributed $H_{ij}$'s with
infinite variance has been investigated in: P. Cizeau, J.-P. Bouchaud,
``Theory of L\'evy matrices'', Phys. Rev.  {\bf E 50} 1810 (1994).} This shows that
$G_{00}$ converges for large $N$ towards a well defined limit
$G_\infty$, which obeys the following limit equation:
\be
\frac{1}{G_\infty(\lambda)} = \lambda - \sigma^2 G_\infty(\lambda). \label{recursionrelation3}
\ee
The solution to this second order equation reads:
\be
G_\infty(\lambda)=\frac{1}{2\sigma^2} \left[\lambda - \sqrt{\lambda^2-4\sigma^2} \right].
\ee
(The correct solution is chosen to recover the right limit for $\sigma=0$.) Now, the only
way for this quantity to have a non zero imaginary part when one adds to $\lambda$ a small
imaginary term $i \epsilon$ which tends to zero is that the square root itself is imaginary.
The final result for the density of eigenvalues is therefore:
\be
\rho(\lambda) = \frac{1}{2\pi \sigma^2} \sqrt{4 \sigma^2 -\lambda^2} \qquad \mbox{for} 
\ |\lambda| \leq 2 \sigma,
\ee
and zero elsewhere. This is the well known `semi-circle' law for the density of states,
first derived by Wigner. This result can be obtained by a variety of other methods
if the distribution of matrix elements is Gaussian.\label{semi-circle} 
\label{bfC}
In finance, one often encounters {\it correlation matrices} $\bf C$,
which have the special property of being positive definite. $\bf C$
can be written as ${\bf C}={\bf H}{\bf H}^\dagger$, where ${\bf
H}^\dagger$ is the matrix transpose of ${\bf H}$.  In general, ${\bf
H}$ is a rectangular matrix of size $M \times N$.  In Chapter 2, $M$
will be the number of asset, and $N$, the number of observation
(days). In the particular case where $N=M$, the eigenvalues of $\bf C$
are simply obtained from those of ${\bf H}$ by squaring them:
\be
\lambda_C = \lambda_H^2.
\ee

If one assumes that the elements of $\bf H$ are random variables, the density of eigenvalues of
$\bf C$ can easily be obtained from:
\be
\rho(\lambda_C) d\lambda_C = \rho(\lambda_H) d\lambda_H,
\ee
which leads to:
\be
\rho(\lambda_C)=
\frac{1}{4\pi \sigma^2} \sqrt{\frac{4 \sigma^2 -\lambda_C}{\lambda_C}} \qquad \mbox{for} 
\ 0 \leq \lambda_C \leq 4 \sigma^2,
\ee
and zero elsewhere. For $N \neq M$, a similar formula exists, which
we shall use in the following. In the limit $N, M \to \infty$, with a
fixed ratio $Q=N/M \geq 1$, one has:\footnote{see: A. Edelmann,
``Eigenvalues and condition numbers of random matrices,''
SIAM J. Matrix Anal. Appl. {\bf 9}, 543 (1988).}

\begin{figure}
\centerline{\hbox{
\epsfig{figure=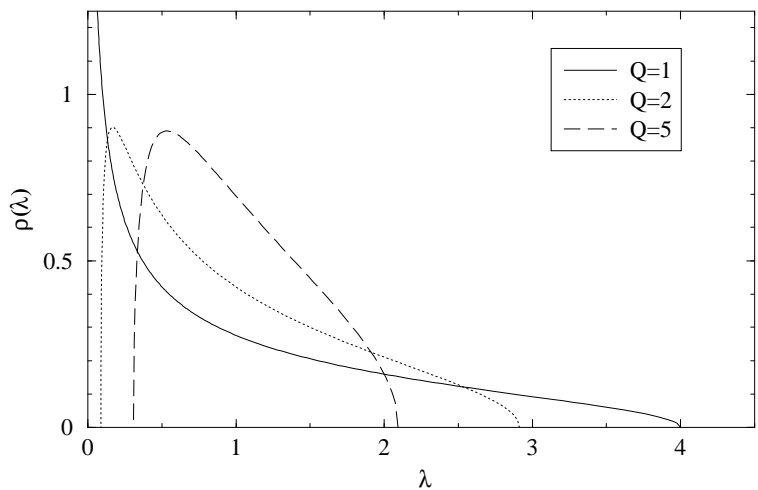,width=10cm}
}}
\caption{Graph of Eq.\ (\protect\ref{rho1}) for Q=1, 2 and 5.}
\end{figure}
\ba\nonumber
\rho(\lambda_C)&=&\frac{Q}{2\pi\sigma^2}
\frac{\sqrt{(\lambda_\subs{max}-\lambda_C)(\lambda_C-\lambda_\subs{min})}}{\lambda},\\
\lambda_\subs{min}^\subs{max}&=&\sigma^2(1+1/Q\pm 2\sqrt{1/Q}),\label{rho1}
\ea
with $\lambda \in 
[\lambda_\subs{min},\lambda_\subs{max}]$. This form is actually also
valid for $Q < 1$, except that there appears a finite fraction of
strictly zero eigenvalues, of weight $1-Q$.

The most important features predicted by Eq.\ (\ref{rho1}) are:
\begin{itemize}
\item The fact that the lower `edge' of the spectrum is strictly positive 
(except for $Q=1$); there is therefore no eigenvalues between 
$0$ and $\lambda_\subs{min}$.  Near this edge, the density of 
eigenvalues exhibits a sharp maximum, except in
the limit $Q=1$ ($\lambda_\subs{min}=0$) where it diverges as 
$\sim {1}/{\sqrt{\lambda}}$.

\item The density of eigenvalues also vanishes above a certain upper edge
$\lambda_\subs{max}$.
\end{itemize}

Note that all the above results are only valid in the limit $N \to
\infty$.  For finite $N$, the singularities present at both edges are
smoothed: the edges become somewhat blurred, with a small probability
of finding eigenvalues above $\lambda_\subs{max}$ and below
$\lambda_\subs{min}$, which goes to zero when $N$ becomes
large.\footnote{see e.g.\ M. J. Bowick, E. Br\'ezin, ``Universal
scaling of the tails of the density of eigenvalues in random matrix
models,'' Phys. Lett {\bf B268}, 21 (1991).}

In Chapter 2, we will compare the empirical distribution of the eigenvalues of the 
correlation matrix of stocks corresponding to different markets with the
theoretical prediction given by Eq.\ (\ref{rho1}).
\section{Appendix A: non-stationarity and anomalous kurtosis}
\label{nonstat3}

In this appendix, we calculate the average kurtosis of the sum  $\sum_{i=1}^N \delta x_i$,
assuming that the $\delta x_i$ can be written as $\sigma_i \epsilon_i$.
The $\sigma_i$'s are correlated as:
\be
\overline{(D_k - \overline D)(D_\ell - \overline D)} =
{\overline D}^2 g(|\ell - k|) \qquad D_k \propto \sigma_k^2. \label{gD}
\ee
Let us first compute $\overline{\left\langle \left(\sum_{i=1}^N \delta
x_i\right)^4 \right\rangle}$, where
$\langle ... \rangle$ means an average over the  $\epsilon_i$'s and the overline
means an average over the  $\sigma_i$'s. If
$\langle \epsilon_i \rangle=0$, and $\langle \epsilon_i  \epsilon_j\rangle=0$ for $i \neq j$, one
finds:
\ba \nonumber
& & \overline{\left\langle \left(\sum_{i,j,k,l=1}^N \delta x_i \delta x_j
\delta x_k \delta x_l\right) \right\rangle} =
\sum_{i=1}^N \overline{\langle \delta x_i^4\rangle}
+ 3 \sum_{i\neq j=1}^N  \overline{\langle \delta x_i^2\rangle\langle
\delta x_j^2\rangle} \\
& & = (3+\kappa_0) \sum_{i=1}^N \overline{\langle \delta x_i^2\rangle^2}
+ 3 \sum_{i\neq j=1}^N
\overline{\langle \delta x_i^2\rangle\langle \delta x_j^2\rangle},
\ea
where we have used the definition of $\kappa_0$ (the kurtosis of $\epsilon$).
On the other hand, one must estimate
$\overline{\left\langle
 \left(\sum_{i=1}^N \delta x_i\right)^2
 \right\rangle}^2$. One finds:
\be
\overline{\left\langle \left(\sum_{i=1}^N \delta x_i\right)^2 \right\rangle}^2
=
\sum_{i,j=1}^N  \overline{\langle \delta x_i^2\rangle} \
\overline{\langle \delta x_j^2\rangle}.
\ee
Gathering the different terms and using the definition (\ref{gD}), one finally establishes
the following general relation:
\be
{\kappa}_N = \frac{1}{N^2 \overline{D}^2}
\left[ N \overline{D}^2 (3+\kappa_0)(1+g(0)) - 3 N \overline{D}^2
+ 3 \overline{D}^2 \sum_{i \neq j=1}^N g(|i-j|)\right],
\ee
or:
\be
\kappa_N = \frac{1}{N} \left[ \kappa_0 +
(3+\kappa_0) g(0)+ 6 \sum_{\ell=1}^{N} (1-\frac{\ell}{N})
g(\ell)\right].
\ee

\section{References}

\begin{myref}{Introduction to probabilities}
\item {\sc W. Feller}, {\it An Introduction to Probability Theory and its Applications},
Wiley, New York, 1971.
\end{myref}

\begin{myref}{Extreme value statistics}
\item{\sc E. J. Gumbel}, {\em Statistics of Extremes\/}, Columbia University
Press, 1958.
\end{myref}

\begin{myref}{Sum of Random Variables, L\'evy distributions}
\item{\sc B. V. Gnedenko} \& {\sc A. N. Kolmogorov}, {\em Limit Distributions
for Sums of Independent Random Variables\/}, Addison Wesley, 1954.

\item{\sc P. L\'evy}, {\em Th\'eorie de l'addition des variables al\'eatoires\/}, Gauthier
Villars, 1937--1954.

\item{\sc G. Samorodnitsky} \& {\sc M. S. Taqqu}, {\it Stable Non-gaussian 
Random Processes}, Chapman \& Hall, New York, 1994. 
\end{myref}

\begin{myref}{Broad distributions in Natural Sciences and in Finance} 
\item{\sc B. B. Mandelbrot}, {\em The Fractal Geometry of Nature\/},
Freeman, 1982.

\item{\sc B. B. Mandelbrot}, {\em Fractals and Scaling in Finance},
Springer, 1997.
\end{myref}

\begin{myref}{Broad distributions in Physics}
\item{\sloppy{\sc E. Montroll}, {\sc M. Shlesinger},
{\it The Wonderful World of Random Walks},
in Nonequilibrium phenomena II, {\em From stochastics to hydrodynamics},
Studies in statistical mechanics XI ({\sc J. L. Lebowitz} \& {\sc E. W. 
Montroll}, eds.) North Holland, Amsterdam, 1984.

}
\item{\sc J.-P. Bouchaud}, {\sc A. Georges},``Anomalous diffusion in 
disordered media: statistical mechanisms, models and physical
applications,'' {\it Physics Reports}, {\bf 195}, 127 (1990).

\item{{\sc G. Zaslavsky}, {\sc U. Frisch} \& {\sc M. Shlesinger}}
{\it L\'evy Flights and Related Topics in Physics}, in
Lecture Notes in Physics {\bf 450}, Springer, 1995.
\end{myref}

\begin{myref}{Theory of Random Matrices}
\item{\sc M. L. Mehta },
{\it Random matrix theory}, Academic Press, New York, 1995.

\item{\sc O. Bohigas, M. J. Giannoni},
{\it Mathematical and Computational Methods in Nuclear Physics}, 
Lecture Notes in Physics, vol. 209, Springer-Verlag, 1983.

\end{myref}

\end{document}